\documentclass[longauth]{aa} 

\usepackage{graphicx}
\usepackage{txfonts}
\usepackage{soul}
\usepackage{natbib}
\usepackage{pdfpages}
\usepackage[colorlinks,urlcolor=cyan,citecolor=blue,linkcolor=blue]{hyperref}
\usepackage{amssymb,amsmath}
\usepackage{array}
\usepackage{booktabs}
\providecommand{\bjdtdb}{\ensuremath{\rm {BJD_{TDB}}}}

\providecommand{\fave}{\langle F \rangle}
\providecommand{\msun}{\ensuremath{\rm\,M_\odot}}
\providecommand{\rsun}{\ensuremath{\rm\,R_\odot}}
\providecommand{\lsun}{\ensuremath{\rm\,L_\odot}}

\providecommand{\me}{\ensuremath{\rm\,M_\oplus}}
\providecommand{\re}{\ensuremath{\rm\,R_\oplus}}

\providecommand{\gaia}{\textit{Gaia}}
\providecommand{\JWST}{JWST}
\providecommand{\TRICERATOPS}{\texttt{TRICERATOPS}}

\begin{document}

\title{TESS discovery of two super-Earths orbiting the M-dwarf stars TOI-6002 and TOI-5713 near the radius valley}

\author{M.~Ghachoui      \inst{\ref{liege},\ref{oukaimeden}}
\and B.V.~Rackham        \inst{\ref{mit_eaps}, \ref{mit_kavli}}                      
\and M.~D\'evora-Pajares \inst{\ref{dpto}, \ref{avature}}                            
\and J.~Chouqar          \inst{\ref{liege}} 
\and M.~Timmermans       \inst{\ref{liege}}                                          
\and L.~Kaltenegger      \inst{\ref{cornell}}                                        
\and D.~Sebastian        \inst{\ref{ubirm}}                                          
\and F.J.~Pozuelos       \inst{\ref{iaa}} 
\and J.D.~Eastman        \inst{\ref{Harvard&Smithsonian}}                            
\and A.J.~Burgasser      \inst{\ref{ucsd}}                                           
\and F.~Murgas           \inst{\ref{IAC_Laguna},\ref{la_laguna}}                     
\and K.G.~Stassun        \inst{\ref{vandy}}                                          
\and M.~Gillon           \inst{\ref{liege}}                                          
\and Z.~Benkhaldoun      \inst{\ref{oukaimeden}}                                     
\and E.~Palle            \inst{\ref{IAC_Laguna},\ref{la_laguna}}                     
\and L.~Delrez           \inst{\ref{liege},\ref{liege_star}}                         
\and J.M.~Jenkins        \inst{\ref{nasa_ames}}                                      
\and K.~Barkaoui         \inst{\ref{liege},\ref{mit_eaps},\ref{IAC_Laguna}}                 
\and N.~Narita           \inst{\ref{Komaba_Tokyo},\ref{Osawa_Tokyo},\ref{IAC_Laguna}} 
\and J.~P. de Leon       \inst{\ref{Komaba_Tokyo1}}                                  
\and M.~Mori             \inst{\ref{Osawa_Tokyo},\ref{Mitaka}}                       
\and A.~Shporer          \inst{\ref{DP&KIA&SR}}                                      
\and P.~Rowden           \inst{\ref{Burlington}}                                     
\and V.~Kostov           \inst{\ref{Greenbelt}}                                      
\and G.~F{\H u}r{\' e}sz \inst{\ref{mit_kavli}}                                      
\and K.A.~Collins        \inst{\ref{Harvard&Smithsonian}}                            
\and R.P.~Schwarz        \inst{\ref{Harvard&Smithsonian}}                            
\and D.~Charbonneau      \inst{\ref{Harvard&Smithsonian}}                            
\and N.M.~Guerrero       \inst{\ref{COPP},\ref{Bryant}}                             
\and G. Ricker           \inst{\ref{mit_kavli},\ref{DP_MIT}}                         
\and E.~Jehin            \inst{\ref{liege_star}}                                     
\and A.~Fukui            \inst{\ref{Komaba_Tokyo},\ref{IAC_Laguna}}                   
\and Y.~Kawai            \inst{\ref{Komaba_Tokyo1}}                                  
\and Y.~Hayashi           \inst{\ref{Komaba_Tokyo1}}                                  
\and E.~Esparza-Borges   \inst{\ref{IAC_Laguna},\ref{la_laguna}}                     
\and H.~Parviainen       \inst{\ref{la_laguna},\ref{IAC_Laguna}}                     
\and C.A.~Clark          \inst{\ref{JPL},\ref{NExScI}}                               
\and D.R.~Ciardi         \inst{\ref{NExScI}}                                         
\and A.S. Polanski       \inst{\ref{Kansas_Lawrence}}                                
\and J. Schleider        \inst{\ref{Greenbelt}}                                      
\and E.A. Gilbert       \inst{\ref{JPL}}                                            
\and I. J.M.\ Crossfield \inst{\ref{Kansas_Lawrence}}                             
\and T. Barclay          \inst{\ref{Greenbelt}}                                   
\and C.D. Dressing       \inst{\ref{ucb}}                                         
\and P.R.~Karpoor        \inst{\ref{ucsd}}      
\and E.~Softich          \inst{\ref{ucsd}}      
\and R. Gerasimov        \inst{\ref{ucsd2},\ref{notredame}}    
\and F.~Davoudi          \inst{\ref{liege}}                                          
}

\institute{Astrobiology Research Unit, Universit\'e de Li\`ege, All\'ee du 6 Ao\^ut 19C, B-4000 Li\`ege, Belgium \label{liege}
\and Oukaimeden Observatory, High Energy Physics and Astrophysics Laboratory, Cadi Ayyad University, Marrakech, Morocco \label{oukaimeden}
\and Department of Earth, Atmospheric and Planetary Science, Massachusetts Institute of Technology, 77 Massachusetts Avenue, Cambridge, MA 02139, USA \label{mit_eaps}
\and Kavli Institute for Astrophysics and Space Research, Massachusetts Institute of Technology, Cambridge, MA 02139, USA \label{mit_kavli}
\and Dpto. Física Teórica y del Cosmos, Universidad de Granada, 18071, Granada, Spain \label{dpto}
\and Avature Machine Learning, Spain\label{avature}
\and Carl Sagan Institute at Cornell University, Space Science Building, Ithaca, NY 14850, USA \label{cornell}
\and School of Physics \& Astronomy, University of Birmingham, Edgbaston, Birmimgham B15 2TT, UK \label{ubirm}
\and Instituto de Astrof\'isica de Andaluc\'ia (IAA-CSIC), Glorieta de la Astronom\'ia s/n, 18008 Granada, Spain \label{iaa}
\and Center for Astrophysics \textbar \ Harvard \& Smithsonian, 60 Garden Street, Cambridge, MA 02138, USA \label{Harvard&Smithsonian}
\and Department of Astronomy \& Astrophysics, University of California San Diego, La Jolla, CA 92093, USA \label{ucsd}
\and Instituto de Astrof\'isica de Canarias (IAC), Calle V\'ia L\'actea s/n, 38200, La Laguna, Tenerife, Spain \label{IAC_Laguna}
\and Departamento de Astrof\'{i}sica, Universidad de La Laguna (ULL), 38206 La Laguna, Tenerife, Spain \label{la_laguna}
\and Department of Physics and Astronomy, Vanderbilt University, Nashville, TN 37235, USA \label{vandy}
\and Space Sciences, Technologies and Astrophysics Research (STAR) Institute, Universit\'e de Li\`ege, All\'e du 6 Ao\^ut 19C, B-4000 Li\`ege, Belgium \label{liege_star}
\and NASA Ames Research Center, Moffett Field, CA 94035, USA \label{nasa_ames}
\and Komaba Institute for Science, The University of Tokyo, 3-8-1 Komaba, Meguro, Tokyo 153-8902, Japan \label{Komaba_Tokyo}
\and Astrobiology Center, 2-21-1 Osawa, Mitaka, Tokyo 181-8588, Japan \label{Osawa_Tokyo}
\and Department of Multi-Disciplinary Sciences, Graduate School of Arts and Sciences, The University of Tokyo, 3-8-1 Komaba, Meguro, Tokyo 153-8902, Japan \label{Komaba_Tokyo1}
\and National Astronomical Observatory of Japan, 2-21-1 Osawa, Mitaka, Tokyo 181-8588, Japan \label{Mitaka}
\and Department of Physics and Kavli Institute for Astrophysics and Space Research, Massachusetts Institute of Technology, Cambridge, MA 02139, USA \label{DP&KIA&SR}
\and Royal Astronomical Society, Burlington House, Piccadilly, London W1J 0BQ, UK \label{Burlington}
\and NASA Goddard Space Flight Center, 8800 Greenbelt Rd, Greenbelt, MD 20771, USA \label{Greenbelt}
\and Crow Observatory, Portalegre, Portugal \label{COPP}
\and Bryant Space Science Center, Department of Astronomy, University
of Florida, Gainesville, FL 32611, USA \label{Bryant}
\and Department of Physics, Massachusetts Institute of Technology, Cambridge, MA 02139, USA \label{DP_MIT}
\and Jet Propulsion Laboratory, California Institute of Technology, Pasadena, CA 91109 USA \label{JPL}
\and NASA Exoplanet Science Institute, IPAC, California Institute of Technology, Pasadena, CA 91125 USA \label{NExScI}
\and Department of Physics and Astronomy, University of Kansas, Lawrence, KS 66045, USA \label{Kansas_Lawrence}
\and Department of Astronomy, University of California Berkeley, Berkeley, CA 94720, USA \label{ucb}
\and Department of Physics, Center for Astrophysics \& Space Sciences, University of California San Diego, La Jolla, CA 92093, USA\label{ucsd2}
\and Department of Physics and Astronomy, University of Notre Dame, South Bend, IN 46556, USA\label{notredame}
\and AIM, CEA, CNRS, Universit\'e Paris-Saclay, Universit\'e de Paris, F91191 Gif-sur-Yvette, France \label{paris}
\and Paris Region Fellow, Marie Sklodowska-Curie Action\label{paris_region_fellow}
\and Cavendish Laboratory, JJ Thomson Avenue, Cambridge, CB3 0HE, UK \label{cavendish}
}

\titlerunning{TESS discovery of two super-Earths orbiting the M-dwarf stars TOI-6002 and TOI-5713 near the radius valley }\authorrunning{Ghachoui et al.}


\abstract{We present the validation of two TESS super-Earth candidates transiting the mid-M dwarfs TOI-6002 and TOI-5713 every 10.90 and 10.44 days, respectively. The first star (TOI-6002) is located $32.038\pm0.019$ pc away, with a radius of $0.2409^{+0.0066}_{-0.0065}$ \rsun, a mass of $0.2105^{+0.0049}_{-0.0048}$ \msun, and an effective temperature of $3229^{+77}_{-57}$ K. The second star (TOI-5713) is located $40.946\pm0.032$ pc away, with a radius of $0.2985^{+0.0073}_{-0.0072}$ \rsun, a mass of $0.2653\pm0.0061$ \msun, and an effective temperature of $3225^{+41}_{-40}$ K. We validated the planets using TESS data, ground-based multi-wavelength photometry from many ground-based facilities, as well as high-resolution AO observations from Keck/NIRC2. TOI-6002 b has a radius of $1.65^{+0.22}_{-0.19}$ \re\ and receives $1.77^{+0.16}_{-0.11} S_\oplus$. TOI-5713 b has a radius of $1.77_{-0.11}^{+0.13} \re$ and receives $2.42\pm{0.11} S_\oplus$. Both planets are located near the radius valley and near the inner edge of the habitable zone of their host stars, which makes them intriguing targets for future studies to understand the formation and evolution of small planets around M-dwarf stars.    
}

\keywords{Planets and satellites -- Techniques: photometric -- Methods: numerical}

\maketitle
\section{Introduction}

The Transiting Exoplanets Survey Satellite \citep[TESS;][]{Ricker2015} has significantly increased the sample of exoplanets smaller than Neptune and larger than Earth (broadly called "super-Earths" and "sub-Neptunes") known to orbit nearby stars.
The absence of planets in this size range in our own Solar System limits us to understand how these planets form and evolve, making them compelling targets to investigate. Although it is arguably important to focus on a wide range of stellar types to study exoplanets, M dwarfs provide appealing targets for our study of super-Earths due to their abundance in our galaxy \citep{Henry2019,Reyle2021} and the significantly more favorable planet-to-star size ratio compared to Sun-analog stars. Especially planets at the inner edge of the Habitable Zone (HZ) provide compelling targets because of the transition from hot, potentially habitable rocky worlds to Venus-like worlds \citep{kaltenegger2023hot}. 
Especially planets close to the inner edge of the HZ of M dwarfs, which could arguably show extended secondary atmospheres resulting from a runaway greenhouse effect, thus provide very interesting targets for observations \citep[][]{Kaltenegger2023,Delrez2022,Turbet2019}

The abundance and the distribution of planets with radii below 4 \re\ for orbits shorter than 100 days allows the identification of specific features that can link their abundance to their formation and evolution like the radius valley \citep[see e.g.][]{Fulton_2017,Cloutier_Menour_2020} and the density valley \citep{Luque&palle2022}, based on their radii and densities distributions. The radius valley features a lack in planets ranging from 1.5 to 2 \re. In this regard, some studies argue that rocky super-Earths are a byproduct of atmospheric escape experienced by gaseous sub-Neptunes driven by the energy radiated by the host stars \citep[see e.g.][]{Lopez_2013, Owen_2013, Jin_2014, Chen_2016} and (or) the energy released by their cooling planetary cores \citep[see e.g.][]{Owen_2013,Ginzburg_2016,Ginzburg_2018}. Other studies argue that some planets are formed as rocky super-Earths in a gas-depleted \citep[see e.g.][]{LopezRice2018,Cloutier_Menour_2020,Cherubim2023} or gas-poor environment with thin gaseous envelopes \citep[see e.g.][]{Lee_2021,Cherubim2023}.
Additional theoretical studies predict that the sub-population of sub-Neptunes consists typically of water-rich worlds formed beyond the water ice-line and migrate toward their host stars \citep[se e.g.][]{Venturini2020,Burn2024,Venturini2024}.
The density valley features three peaks in the density distribution defining rocky super-Earths, water-rich worlds, and puffy sub-Neptunes. \cite{Luque&palle2022} suggest that the pebble accretion is the main mechanism forming these planets with super-Earths formed within the ice line and water-rich world formed outside it \citep[se e.g.][]{Venturini2020,Burn2024,Venturini2024}. For now the mechanisms forming super-Earths remain unclear, making observations of planets in this mass-radius regime a critical part to understand it.     

TESS is increasing the number of small exoplanets amenable to mass and atmospheric characterization \citep{Ricker2015,Sullivan2015Ap}. Detailed studies of such exoplanets will allow insights into the origin of the radius and density valleys in the near future. Such analyses should shed further light on the mechanisms dominating the formation and evolution of super-Earths. In addition, the analyses of exoplanets with incident irradiation between Earth and Venus will allow key insights into the limits of the HZ \citep[e.g.][]{kaltenegger2023hot}.

This paper reports the validation of two super-Earths, TOI-6002 b ($1.65^{+0.22}_{-0.19}$ \re) and TOI-5713 b ($1.77_{-0.11}^{+0.13}$ \re) orbiting two mid- type M-dwarf stars, TOI-6002 
and TOI-5713 
. They were first detected by TESS, and validated in this study using various ground-based observations (see Sect.~\ref{obs:phot}). The masses of both planets are left unconstrained by our observations, but their radii suggest rocky or water-rich composition. Note that precise mass measurements should be possible with state-of-the art Doppler spectographs to derive their masses, as discussed in more detail in Sect.~\ref{discus}.

The paper is structured as follows. Sect.~\ref{obs:phot} introduces the data from TESS and all ground-based observations. Stellar characterization, validation of the planetary nature of the two TESS candidates, and global data analyses are covered in Sect.~\ref{anlys}. Results, potential for mass measurements, and atmospheric characterization are discussed for both planets in Sect.~\ref{discus}. Sect.~\ref{concl} summarizes our findings.

\section{Observations} \label{obs:phot}

This section presents all observations of TOI-6002 and TOI-5713 obtained from space and ground-based instruments. In Table \ref{tab:obs}, we summarize all the ground-based, photometric time-series observations, as well as the observation parameters. We note that the differential photometry was performed using uncontaminated circular apertures for all ground-based observations. 

\begin{table*}
\begin{center}
\caption{Ground-based facilities and their corresponding parameters used in the time-series photometric observations of TOI-6002 and TOI-5713.}. \label{tab:obs}
\resizebox{\textwidth}{!}{%
\begin{tabular}{l c c c c c c c c}
\toprule
Date (UT)      & Filter    &  Facility             & Pixel Scale (\arcsec/pixel) & Exposure Time [s] & Number of Exposures & Airmass Range & Aperture Size (\arcsec) & Detrending Parameters\\
\midrule
TOI-6002 \\
\midrule
13 May 2023   & $i'$       & LCO-SINISTRO (1.0 m)  & 0.389 & 86        & 93       & [1.00,1.40] & 2.33 & Airmass+Background \\
\midrule
24 May 2023   & $i'$       & LCO-SINISTRO (1.0 m)  & 0.389 & 86        & 83       & [1.04,1.89] & 2.33 & Airmass+FWHM       \\
\midrule
19 August 2023& $r', z_s$  & LCOGT MuSCAT3 (2.0 m) & 0.265 & 46, 32    & 251, 339 & [1.03,1.30] & 2.38 & Airmass+FWHM       \\
              & $g', i'$   &                       & ...   & 207, 28   & 35, 375  & ...         & ...  & Airmass+Background \\
\midrule
TOI-5713 \\
\midrule
17 June 2023 & $I+z$       & TRAPPIST-North (0.6 m)& 0.640 & 50        & 209      & [1.25,2.88] & 5.40 & Airmass+FWHM       \\
\midrule
08 July 2023 & $r', z_s$   & TCS-MuSCAT2 (1.52 m)  & 0.440 & 50, 25    & 195, 457 & [1.16,1.94] & 6.96 & Airmass+FWHM       \\  
             & $i'$        &                       & ...   & 15        & 234      & ...         & ...  & Airmass+Background \\
\hline 
\end{tabular}%
}
\end{center}
Aperture size column presents the optimum uncontaminated apertures used for extracting the light curves. Detrending parameters were chosen on the basis of maximizing the log-likelihood of each transit.
\end{table*}

\subsection{TESS photometry} \label{TESS:phot}

TOI-6002 was observed by TESS in sectors 14 (18 July to 14 August 2019), 41 (24 July to 20 August 2021), and 54--55 (9 July to 1 September 2022). TOI-5713 was observed by TESS in sectors 16 (12 September to 06 October 2019), 22--23 (19 February to 15 April 2020), and 49 (26 February to 25 March 2022). Note that this target was also observed in sectors 74 and 75 from 03 January to 26 February 2024. These two sectors were not included in our analyses here (see Sect.~\ref{Modeling}), but the data appears consistent with our results and will be used as part of future research on these planets. 

The TESS time-series observations were processed by the TESS Science Processing Operations Center (SPOC) pipeline \citep{Jenkins_2016,Jenkins_2020a}. SPOC calibrates the image data, carries out quality control, labels bad data, and performs simple aperture photometry (SAP) for each star in the field of view. Then, it corrects for instrumental systematic offsets using a Presearch Data Conditioning module \citep[PDC:][]{Smith_2012,Stumpe_2012,Stumpe_2014} and produces pre-search data conditioned light curves. Appendix~\ref{TESS:fov} shows the target pixel files (TPFs) and the SAP apertures used for photometry, along with the overlaid locations of nearby \gaia\  DR2 \citep{gaia_2018} sources. 

The SPOC pipeline executed a transit search of the combined light curve from sectors 14--55 on 11 September 2022 for TOI-6002 and from sectors 16--49 on 05 May 2020 for TOI-5713 \citep{Jenkins_2002,Jenkins_2020a}, yielding a threshold crossing event (TCE, a statistically significant periodic, transit-like signal on the target star) with a 10.9 day period for the first target and a TCE with a 10.44 day period for the second. Initial limb-darkened transit models were fitted \citep{LI_2019PASP} and a suite of diagnostic tests were performed to prove or refute the planetary origin of the signals \citep{Twicken_2018PASP} for both TCEs. The difference image centroiding tests conducted for the sectors 14-55 search of TOI-6002 and the sectors 16-49 search of TOI-5713 constrained the location of the target star to within 4.6$\pm$4.1\arcsec\ and 3.4$\pm$3.2\arcsec\ of the transit source, respectively.  The two transit signatures passed all diagnostic tests presented in the SPOC data validation reports to discard different possible false positives, mainly eclipsing binaries and blended eclipsing binaries. The TESS Science Office (TSO) evaluated the vetting information and issued an alert for TOI-6002 b on 01 December 2022 and for TOI-5713 b on 8 August 2022.

For our analyses, we downloaded the 2-minute PDC-SAP light curves from the Mikulski Archive for Space Telescopes (MAST\footnote{\url{https://archive.stsci.edu/}}). Using the "QUALITY" flag, we identified all the bad-quality data points that were removed from the data files. The year two data of TESS were subject to sky background overestimation in the SPOC pipeline for dim and (or) crowded targets, affecting the transit depth and, thus, the planet radius. TOI-6002 is a crowded-field target and was subjected to this effect in sector 14. We corrected the PDC-SAP light curves of sector 14 using equation 2 described in section 4.2 of the data release notes of sector 27\footnote{\url{https://tasoc.dk/docs/release_notes/tess_sector_27_drn38_v02.pdf}} (see section 4.2, equation 2). This effect was negligible for TOI-5713 due to the low level of crowding. The light curve of TOI-5713 showed some clear flare-like signals, which we removed using an upper 3$\sigma$ clipping.
We used \texttt{wot$\bar{\mathrm{a}}$n} \citep{Hippke2019} to detrend the light curves for stellar variability, utilizing a bi-weight filter with a window 2.5 times longer than each planets' transit duration. The obtained lightcurves are presented in Appendix~\ref{TESS:LC}, with the locations of the transit signals highlighted with different colors.

\subsection{Ground-based photometry}\label{all_TESS}

\subsubsection{LCOGT-1m0 photometry}

We used the 4096$\times$4096 SINISTRO CCD camera on the 1.0-m robotic telescope of the Las Cumbres Observatory Global Telescope (LCOGT; \citealt{Brown_2013}) network node at Teide Observatory (Tenerife, Spain) to observe two full transits of TOI-6002\,b on UT 2023 May 13 and 24.
Table \ref{tab:obs} provides the observation parameters. 
We used the standard LCOGT {\tt BANZAI} pipeline \citep{McCully_2018SPIE10707E} to calibrate the science images, and we used \texttt{AstroImageJ} \citep[AIJ:][]{Karen2017} for the photometric extraction.

\subsubsection{LCOGT MuSCAT3 photometry}

We used the MuSCAT3 multi-band imager \citep{Narita_2020SPIE11447E} on the LCOGT 2m Faulkes Telescope North at Haleakala Observatory on Maui (Hawaii) to observe one full transit of TOI-6002\,b on UTC August 19, 2023 (see Table \ref{tab:obs} for more details).
The multi-band observation covered $g$, $r$, $i$, and $z_{\rm s}$ bands simultaneously. 
The data reduction and aperture photometry were performed in a similar way as described in the previous section.

\subsubsection{TRAPPIST-North photometry}

We used the CCD camera on TRAPPIST-North, a 0.6-m robotic telescope at Oukaimeden Observatory in Morocco (\citealt{jehin2011,gillon2013,Barkaoui2019,Ghachoui2023A&A}), to observe a full transit of TOI-5713\,b on UTC June 17, 2023 (see Table \ref{tab:obs}).
We observed with the non-standard I+$z'$ filter.
The data reduction and photometry were performed using {\tt AstroImageJ}.

\subsubsection{MuSCAT2 photometry}

For TOI-5713\,b, we observed one full transit on UTC July 8, 2023 (see Table \ref{tab:obs}), using the multi-color imager ($g$, $r$, $i$, and $z_{\rm s}$ bands) MuSCAT2 \citep{Narita2019}, mounted on the 1.52-m Telescopio Carlos S\'anchez (TCS) at Teide Observatory, Spain. 
We used the $g$ band images for guiding, as the star is faint in the blue optical ($B \sim 17$ mag). 
We calibrated and reduced the raw data using the code and procedure of \citet{Parviainen2019}. 
Due to the short exposure time and the faintness of the target, the MuSCAT2 $g$ band photometry presented a much larger scatter than in the other bands, and it was not used in the analysis presented below.

We also observed a full transit of TOI-6002\,b with MuSCAT2 on UTC October 01, 2023.
However, this observation was affected by clouds around the mid-transit time, and thus was omitted in the global analyses.

\subsection{Spectroscopic observations}

\subsubsection{IRTF/SpeX}
We used the SpeX spectrograph \citep{Rayner2003} on the 3.2-m NASA Infrared Telescope Facility (IRTF) to collect a  near-infrared spectrum of TOI-5713 on 2023 May 16 (UT) and TOI-6002 on 2023 Oct 31 (UT).
On both nights, conditions were clear with seeing of 0$\farcs$6--0$\farcs$7.
We used the short-wavelength cross-dispersed (SXD) mode, and aligned the $0\farcs3 \times 15''$ slit to the parallactic angle.
The resulting spectra span 0.80–2.42\,$\mu$m and have a resolving power of $R{\sim}2000$.
We collected six integrations of 139.9\,s on TOI-5713 and 299.8\,s on TOI-6002, using an ABBA nod pattern.
Immediately after the science observations, we gathered a standard set of flat-field and arc-lamp calibrations and observed A0\,V standard stars.
We reduced the data with Spextool v4.1 \citep{Cushing2004}, following the standard instructions from the Spextool User's Manual\footnote{Available at \url{http://irtfweb.ifa.hawaii.edu/~spex/observer/}.}.
The final spectrum of TOI-5713 has a median per-pixel S/N of 92 with peak values in the $J$, $H$, and $K$ bands of of 116, 131, and 121, respectively.
For TOI-6002, the final spectrum has a median S/N per pixel of 156 with peak values of 195, 219, and 205 in the $J$, $H$, and $K$ bands, respectively.

\subsubsection{Shane telescope/kast double spectrograph observations}

We observed TOI-5713 and TOI-6002 with the Kast Double Spectrograph \citep{kastspectrograph} on the 3-m Shane Telescope at Lick Observatory. 
TOI-5713 was observed on 2023 July 7 (UT) in clear conditions with 1.1$\arcsec$ seeing; 
TOI-6002 was observed on 2023 November 5 (UT) in partly cloudy conditions with 1$\farcs$2 seeing.
We used both the red and blue channels of Kast split by the D57 dichroic at 5700~{\AA} and the 1.5$\arcsec$-wide slit for both observations. 
In the red channel, we used the 600/7500 grating providing 5900--9000~{\AA} wavelength coverage at an average resolution of $\lambda/\Delta\lambda$ = 1450.
In the blue channel, we used the 600/4310 grism providing 3600--5600~{\AA} wavelength coverage at an average resolution of $\lambda/\Delta\lambda$ = 1100. 
For TOI-5713, we obtained one blue exposure of 1200~s and two red exposures of 600~s each at an average airmass of 1.07; 
for TOI-6002, we obtained one blue exposure of 900~s and two red exposures of 450~s each at an average airmass of 1.03. 
We observed a nearby G2~V star immediately after each source for calibration of telluric absorption, and spectrophotomeric flux standards PG 1708+60 and BD +28 4211 \citep{1992PASP..104..533H,1994PASP..106..566H} during each night for flux calibration.
Flat-field lamps, HeNeAr arc lamps, and HgCd arc lamps were also observed at the start of each night for pixel response and wavelength calibration. 

All data were reduced using the kastredux package\footnote{kastredux:~\url{https://github.com/aburgasser/kastredux}.}, which included image calibration (pixel response correction, conversion to electrons/second, bad pixel masking), boxcar extraction with median background subtraction, wavelength calibration, flux calibration, and correction of telluric absorption in the red channel.
The final spectra have median signals-to-noise of 140 in both red channels (near 7450~{\AA}) and 45 in the blue channel of TOI-6002 (near 5100~{\AA}). Due to calibration issues, we discarded the blue observations of TOI-5713.

\subsection{High-resolution imaging} \label{OBS:AO}

To constrain possible contamination by bound or line-of-sight companions \citep{ciardi2015}, We observed TOI-5713 and TOI-6002 with near-infrared adaptive optics (AO) imaging at the Keck (2023-Jun-10 UT) and at the Palomar (2023-Jun-08 UT) Observatory. 

TOI-5713 was observed with the NIRC2 instrument on Keck-II (10 m) using the natural guide star AO system \citep{wizinowich2000} utilizing a standard 3-point dither pattern. Each position in the dither pattern is offset from the previous position by 3$\arcsec$. The 3-point dither pattern was performed 3 times, each pattern offset from the previous pattern by $0.5\arcsec$, resulting in a total of 9 on-target exposures. NIRC2 was used in the narrow-angle mode with the full field of view ($\sim10\arcsec$) and a pixel scale of roughly $0.0099442\arcsec$ per pixel. Observations were made in the Kcont filter $(\lambda_o = 2.2706; \Delta\lambda = 0.0296~\mu$m) with an integration time of 15 seconds per dither position, yielding a total of 135 seconds on-source time.
	
We observed TOI-6002 with the PHARO instrument \citep{hayward2001} on the Palomar Hale (5m) using the P3K natural guide star AO system \citep{dekany2013} in the narrowband Br-$\gamma$ filter $(\lambda_o = 2.1686; \Delta\lambda = 0.0326~\mu$m). The PHARO pixel scale is $0.025\arcsec$ per pixel. We twice repeated a standard 5-point quincunx dither pattern with steps of 5\arcsec, separating each repeat by 0.5\arcsec. We used a single frame integration time of 15 sec, yielding a total integration of 225 seconds.
    
Flat fields were taken on-sky, dark-subtracted, and median averaged. We generated sky frames using the median average of the dithered science frames. We then sky-subtracted and flat-fielded each science frame and combined the calibrated science frames into a single mosaiced image.  The final resolutions of the combined dithers, determined  from the full-width half-maximum (FWHM) of the point spread functions, were 0.057\arcsec\ and 0.108\arcsec\ for TOI-5713 and TOI-6002, respectively.  
	
\begin{figure*}
	\centering
	\includegraphics[scale=0.09]{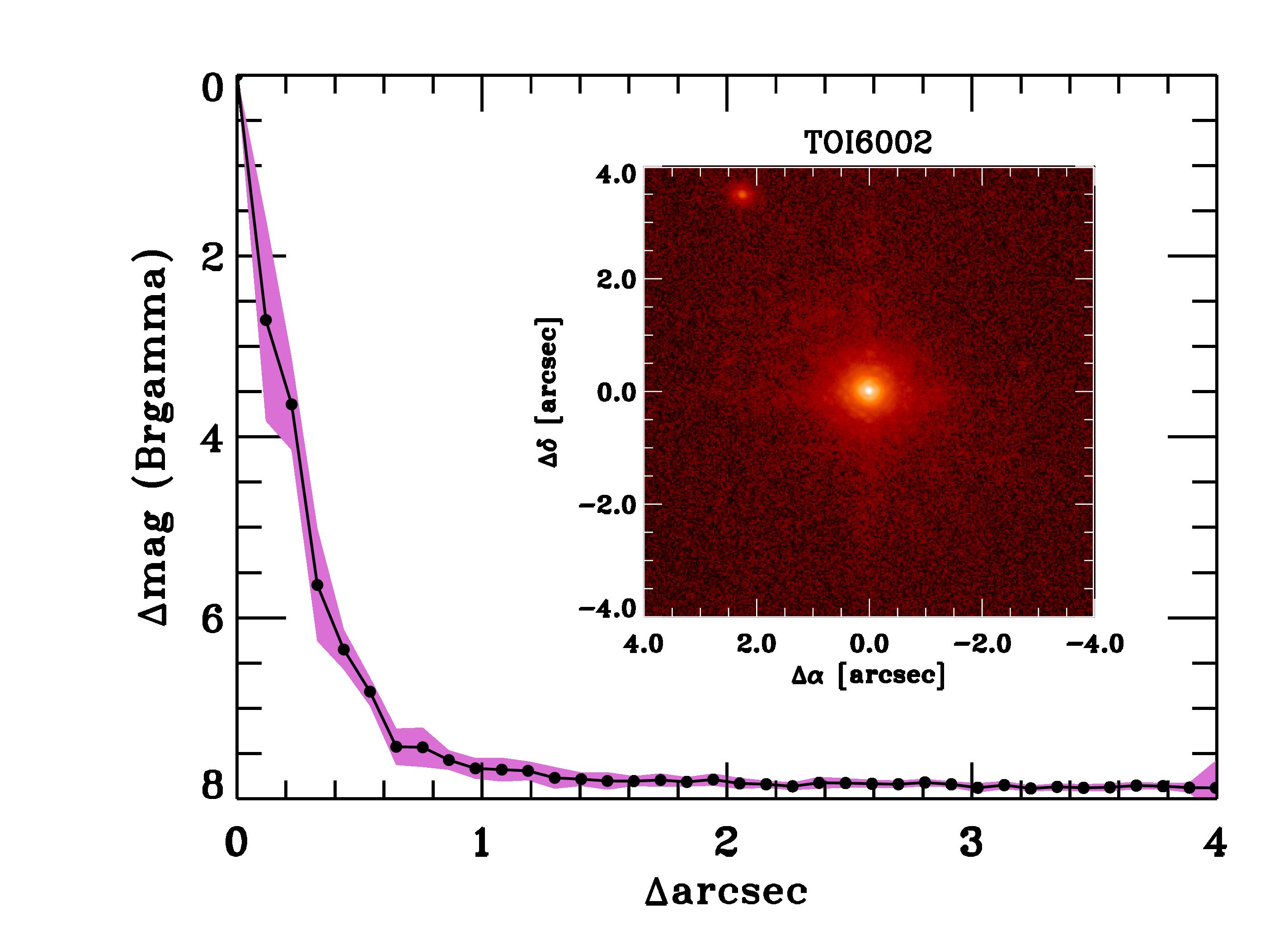}
	\includegraphics[scale=0.09]{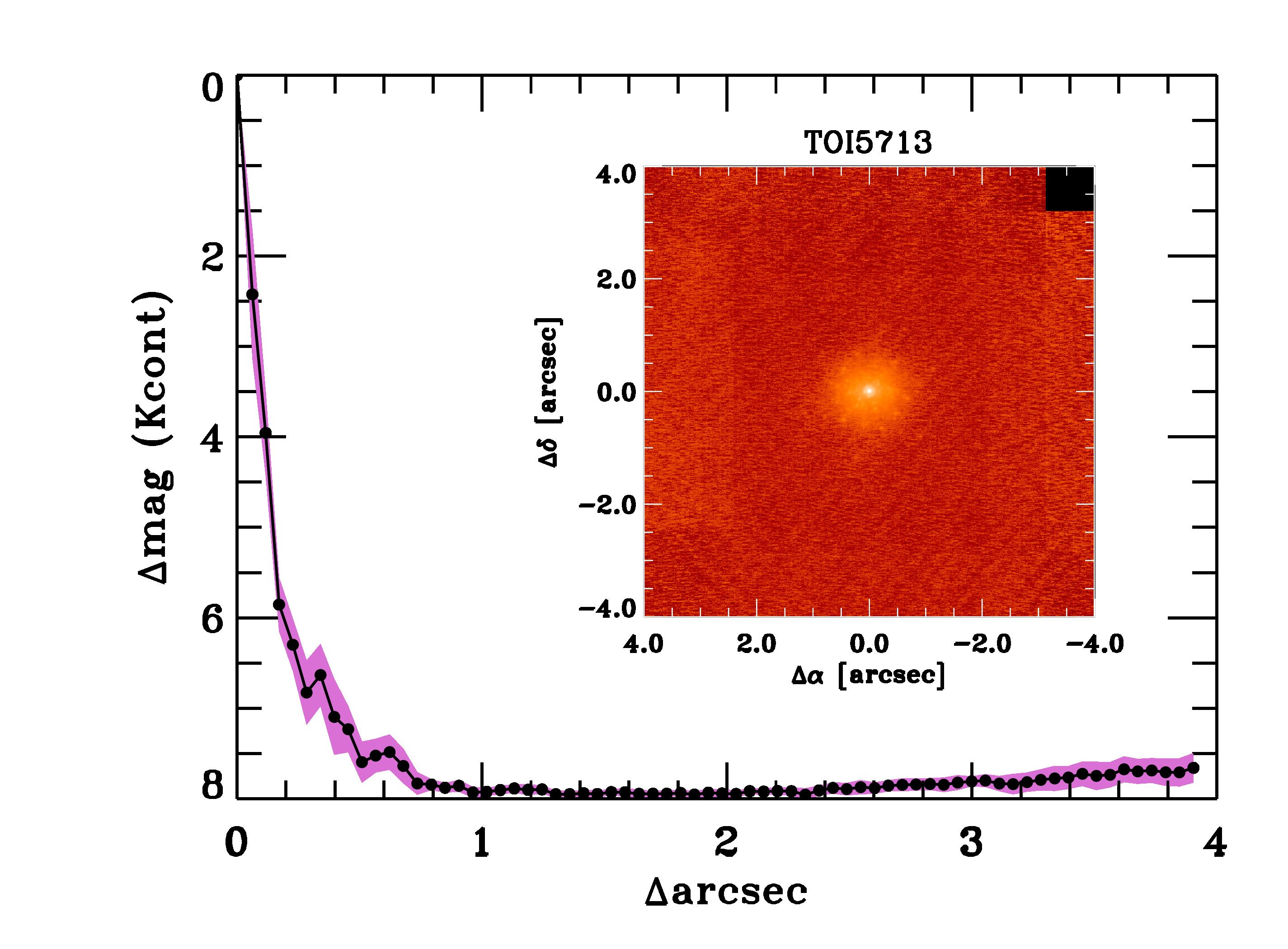}
	\caption{NIR AO imaging and sensitivity curve for TOI~5713 (Keck) and TOI~6002 (Palomar). {\it Inset:} Images of the central portion of the data.}
	\label{fig:ao_images}
\end{figure*}

We assessed the sensitivities of the final combined AO image by injecting simulated sources azimuthally around the primary target, following the approach of \citet{furlan2017}.
Fig.~\ref{fig:ao_images} shows the resulting sensitivity curves. 
We do not detect close-in ($\lesssim 1\arcsec$) stellar companions for either star. 
The Keck and Palomar data have sensitivities of $\delta mag \approx 7.5$ mag and $6.9$ mag at 0.5\arcsec, respectively. 
This corresponds to a spatial limit $\lesssim 32$\,au for TOI-6002 and $\lesssim 40.94$\, au for TOI-5713. 

\subsection{\gaia\ assessment}
\label{sub:gaia}

In addition to high-resolution imaging, we used \gaia\ to pinpoint any widely separated, bound stellar companions.
The \gaia\ catalog does not identify any nearby sources that have the same distance and proper motion as TOI-5713 or TOI-6002 \citep[see also][]{mugrauer2020,mugrauer2021}.  
There is a 4\arcsec\ neighbor source to TOI-6002, but it is a line-of-sight source at $>2000$ pc and unrelated to TOI-6002.
     
We also used the \gaia{} renormalized unit weight error (RUWE) to assess the possibility of any close-in companions.
RUWE values $\gtrsim 1.4$ indicate excess astrometric noise, which can potentially be caused by a bound companion \citep{ziegler2020}. 
The \gaia\ DR3 RUWE values for TOI-5713 and TOI-6002 are 1.3 and 1.1, respectively, indicating astrometric solutions that are consistent with single-star models. 

\section{Analyses} \label{anlys}

\subsection{Stellar characterization} \label{star}

\subsubsection{Spectroscopic analysis}\label{sec:star_spec}
\begin{figure}
    \centering
    \includegraphics[width=\columnwidth, height=0.5\textheight, keepaspectratio]{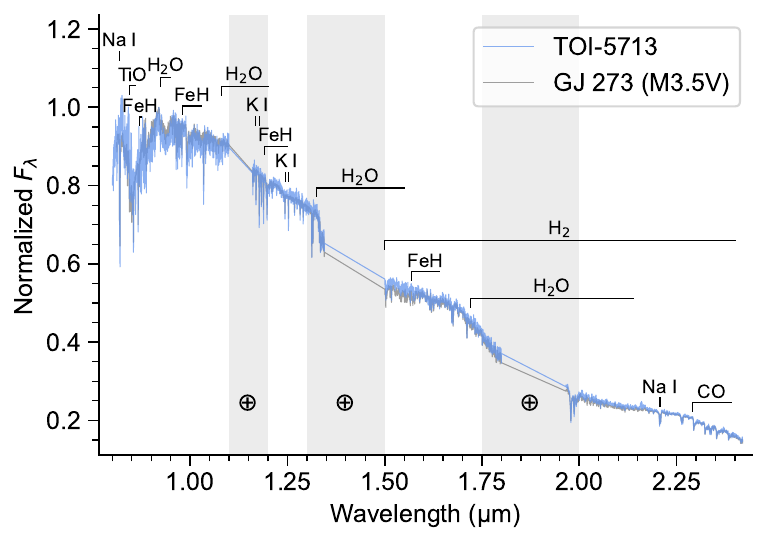}
    \includegraphics[width=\columnwidth, height=0.5\textheight, keepaspectratio]{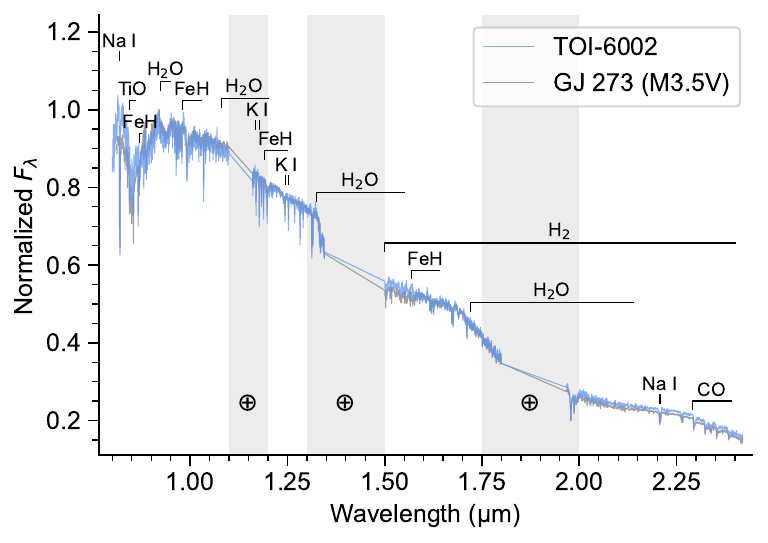}
    \caption{
        SpeX SXD spectra of TOI-5713 (top) and TOI-6002 (bottom).
        Spectra of the targets (blue) and the M3.5V spectral standard GJ 273 (grey) are shown.
        Annotations indicate strong spectral features of M dwarfs, and shaded regions cover wavelengths with strong telluric absorption.
}
    \label{fig:spex}
\end{figure}

Fig.~\ref{fig:spex} shows the SpeX SXD spectra of TOI-5713 and TOI-6002.
To assign spectral types, we compared the spectra to single-star spectral standards in the IRTF Spectral Library \citep{Cushing2005, Rayner2009} using the SpeX Prism Library Analysis Toolkit \citep[SPLAT, ][]{splat}.
For both stars, we find the single best match to the M3.5 dwarf GJ\,273, with M3 and M4 dwarf standards providing the next closest matches.
Accordingly, we adopt an infrared spectral type of M3.5 $\pm$ 0.5 for both targets.
Following \citet{Delrez2022}, we used the \citet{Mann2013} relation between the equivalent widths of the $K$-band Na\,\textsc{i} and Ca\,\textsc{i} doublets, the H2O--K2 index \citep{Rojas-Ayala2012}, and stellar metallicity to estimate an iron abundance of $\mathrm{[Fe/H]} = -0.03 \pm 0.13$ for TOI-5713 and $\mathrm{[Fe/H]} = -0.18 \pm 0.12$ for TOI-6002.

\begin{figure}
    \centering
    \includegraphics[width=\columnwidth, keepaspectratio]{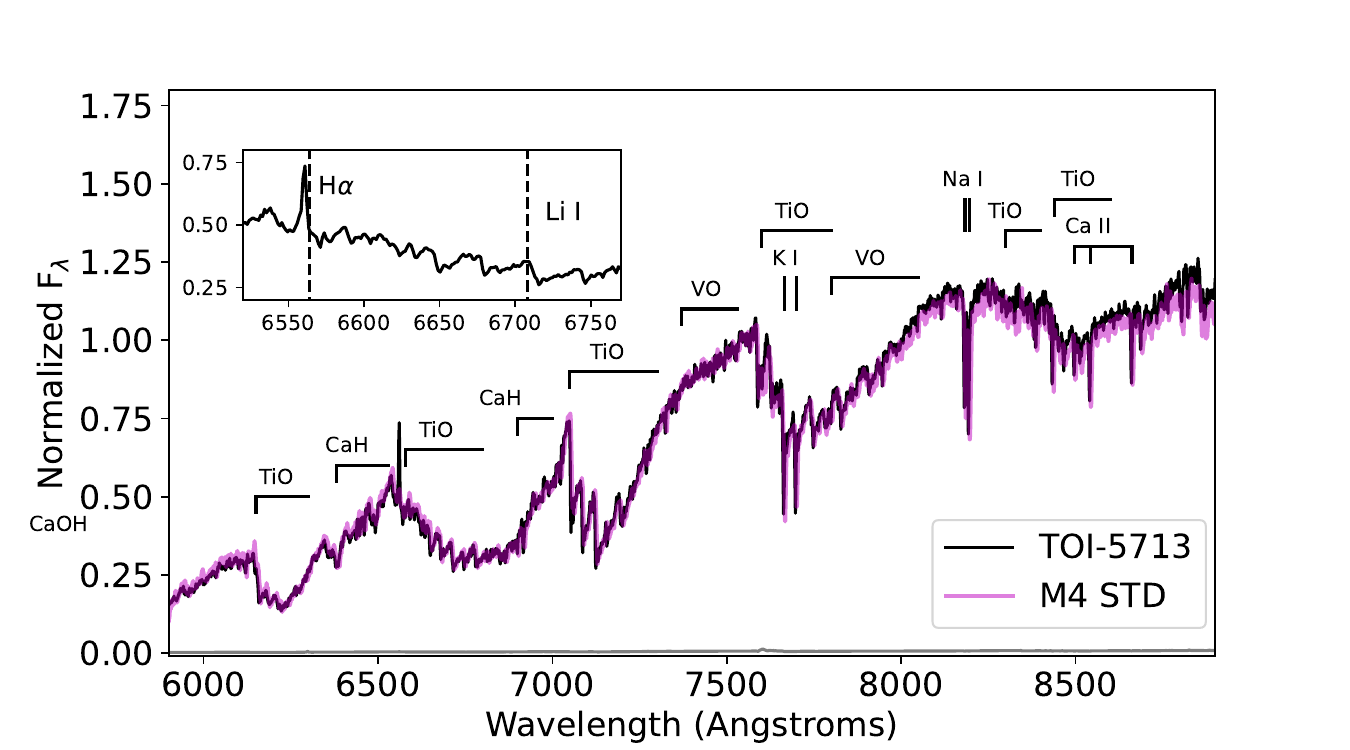}
    \includegraphics[width=\columnwidth, keepaspectratio]{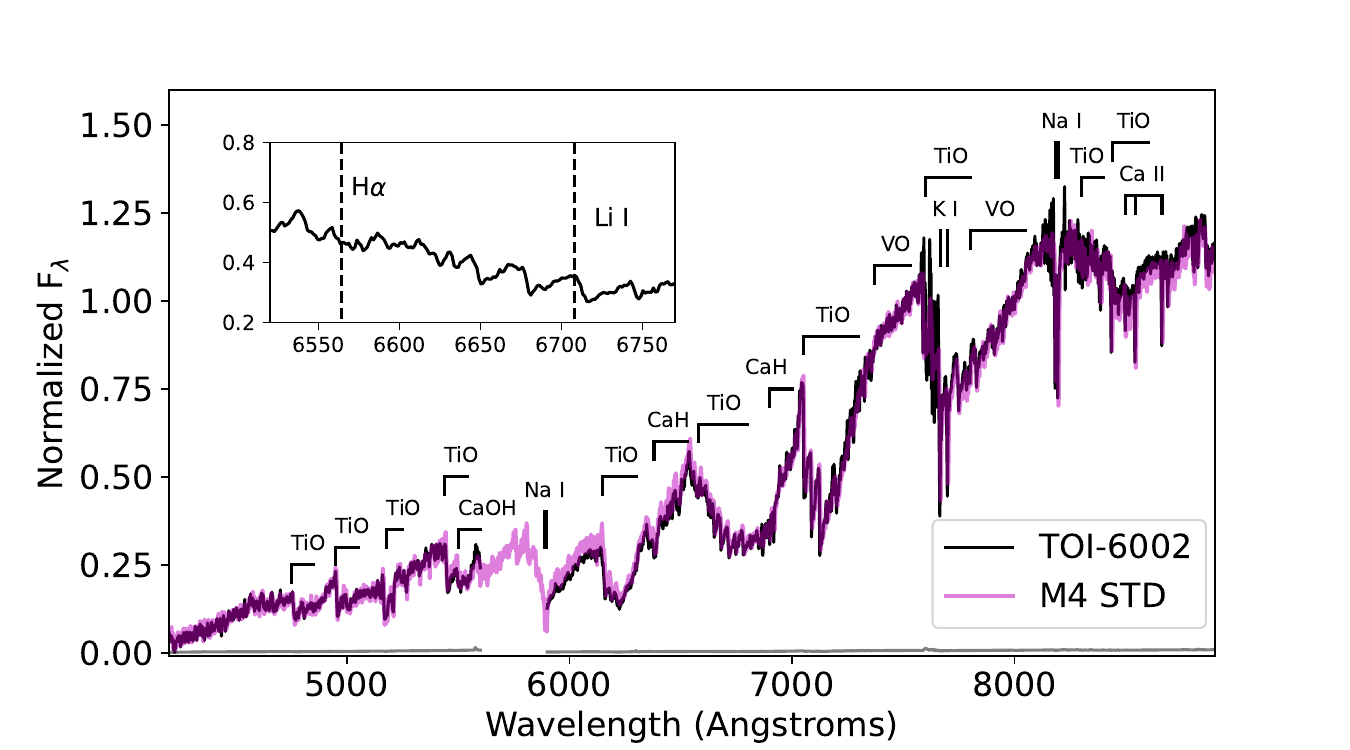}
    \caption{
        Shane/Kast blue and red optical spectra (black lines) of TOI-5713 (top) and TOI-6002 (bottom) compared to their best-fit M4 SDSS standard templates (magenta line; \citealt{2007AJ....133..531B}).
        Data are normalized at 7400~{\AA}, and the gap between 5600~{\AA} and 5900~{\AA} in the spectrum of TOI-6002 corresponds to the dichroic split between the Kast blue and red channels.
        Key spectral features across the 4200--8900~{\AA} region are labeled. Inset boxes highlight the regions around the 6563~{\AA} H$\alpha$ and 6708~{\AA} Li~I lines.
}
    \label{fig:kast}
\end{figure}

The Shane/Kast optical spectra of TOI-5713 and TOI-6002 are shown in Fig.\,\ref{fig:kast}. Comparison to SDSS M dwarf templates from \citet{2007AJ....133..531B} shows a best-match to the M4 template in both cases, 
with index-based classifications from \citet{1995AJ....110.1838R}; \citet{1997AJ....113..806G}; and \citet{2003AJ....125.1598L} indicating equivalent classifications of M3.5--M4, also consistent with the near-infrared classifications.
We detect H$\alpha$ emission in TOI-5713 with an equivalent width of 1.94$\pm$0.06~{\AA}, corresponding to $\log_{10}{L_{H\alpha}/L_{bol}}$ = -4.16$\pm$0.07 based on the $\chi$ factor relation of \citet{2014ApJ...795..161D}. For TOI-6002, we see no evidence of H~I emission, with an H$\alpha$ equivalent width limit of 0.6~{\AA}. These emission signatures indicate ages less than and greater than $\sim$4.5~Gyr, respectively \citep{2008AJ....135..785W}. 
We measured metallicity-sensitive $\zeta$ indices \citep{2007ApJ...669.1235L,2013AJ....145...52M} of 1.07$\pm$0.01 and 1.02$\pm$0.01 for TOI-5713 and TOI-6002, respectively, consistent with near-solar metallicities of [Fe/H] = +0.08$\pm$0.20 and [Fe/H] = +0.04$\pm$0.20 \citep{2013AJ....145...52M}, again formally consistent with the metallicities inferred from the near-infrared spectra.

\subsubsection{SED and empirical relations within the global modeling}
\label{JointSED}

\begin{figure*}[h!] 
    \centering
    \includegraphics[width=0.49\linewidth,trim=10 10 10 5,clip]{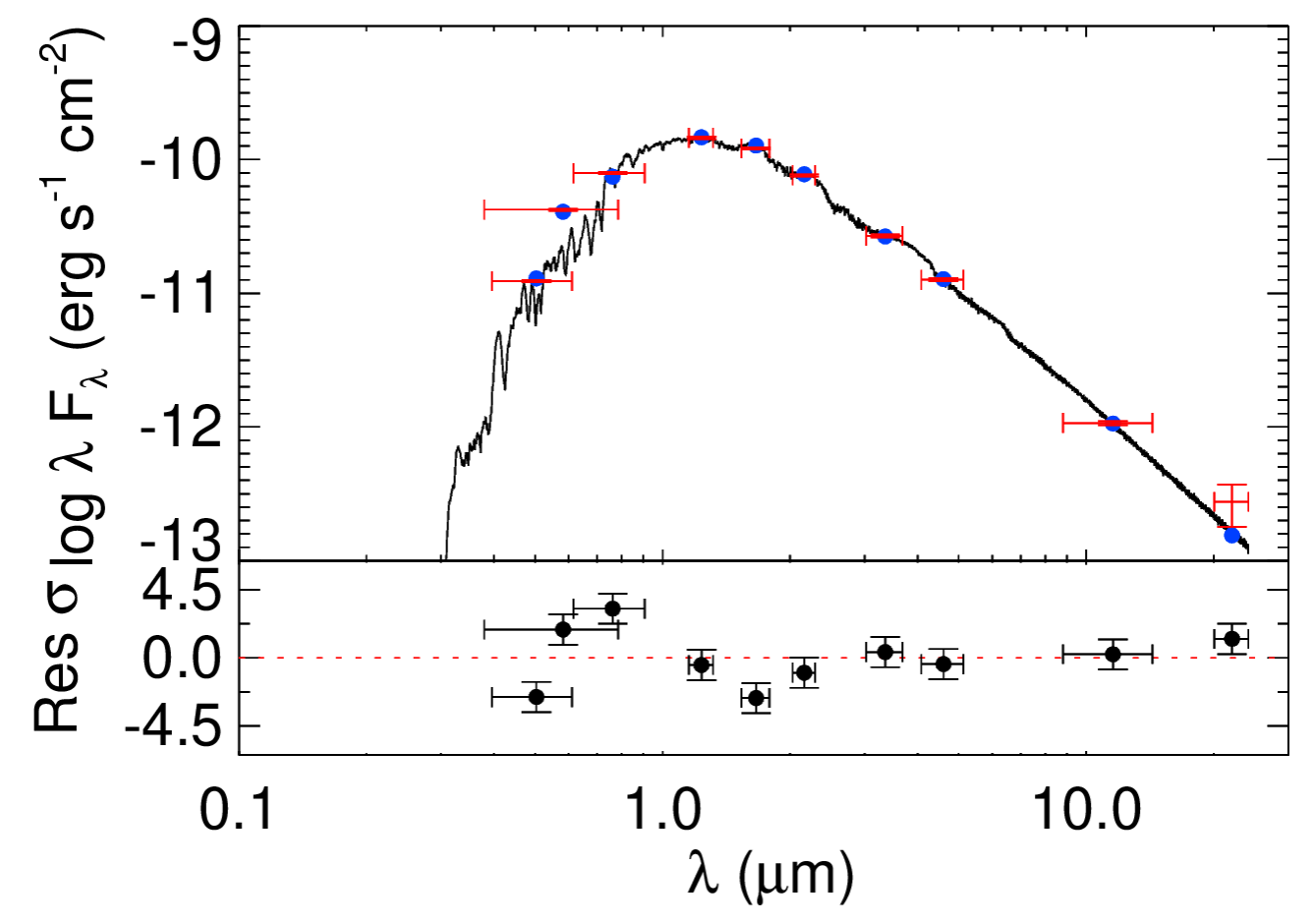}
    \includegraphics[width=0.49\linewidth,trim=10 10 10 5,clip]{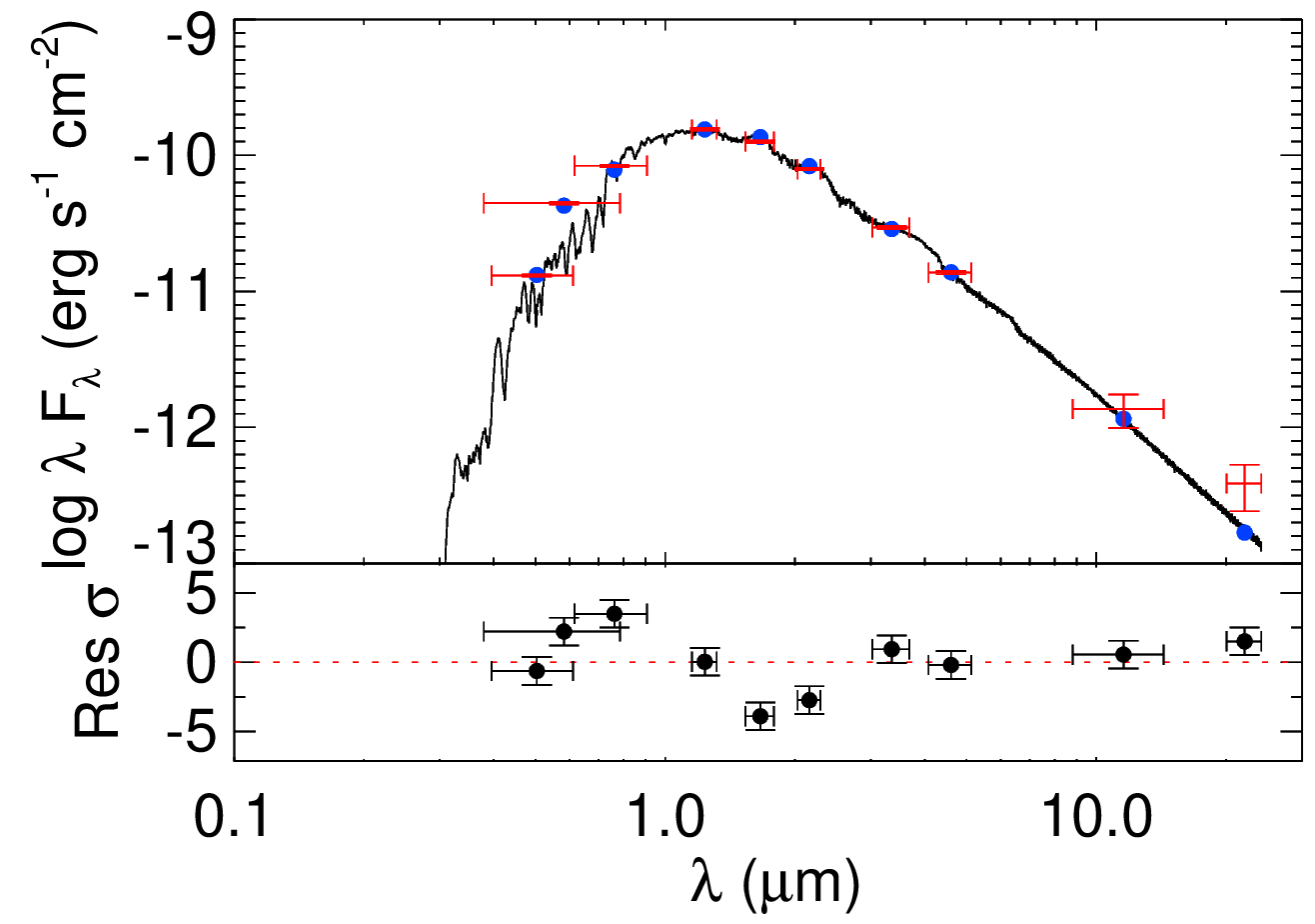}
\caption{Spectral energy distribution (SED) of TOI-5713 (left) and TOI-6002 (right). Black lines represent the SED best-fit models. Blue circles represent the average broadband flux measurements based on the SED model. Horizontal error bars show the filters bandwidth, and the vertical error bars show the uncertainty on the flux measurements. \label{fig:sed}}
\end{figure*}

We performed a Spectral Energy Distribution (SED) fit jointly with transit modeling (see Sect.~\ref{Modeling}) using the latest version of \texttt{EXOFASTv2}\footnote{\url{https://github.com/jdeast/EXOFASTv2/branches}} \citep{Eastman_2019} software package. The latest version of \texttt{EXOFASTv2} 
comes with the possibility of simultaneously utilizing the empirical relations of \cite{Mann_2015} to determine the stellar radius $R_*$, and of \cite{Mann_2019} to determine the stellar mass $M_*$. We activated this feature using the two corresponding flags, \texttt{MANRAD} and \texttt{MANMASS}. The $R_*$ and $M_*$ values obtained by \texttt{EXOFASTv2} are optimized and self-consistent compromises between the constraints from the SED, empirical relations, and transits (via stellar density). The stellar parameters obtained in this case are more precise than when using the MIST stellar evolution models \citep{Dotter_2016}, which was disabled by setting the \texttt{NOMIST} flag as recommended in \cite{Eastman_2019}. For the SED fit, we used the $JHK_S$ magnitudes from {\it 2MASS}, the W1--W4 magnitudes from {\it WISE} \citep{Cutri_2013}, the $G_{\rm BP} G_{\rm RP}$ magnitudes from \gaia\ DR3 \citep{GaiaCollaboration_2020}, and the {\it Pan-STARRS\/} \citep{Chambers+2016} $yz$ magnitudes (see Table \ref{table:spec+astro}). We applied a Gaussian prior on the \gaia\ EDR3 parallax, which we corrected for systematics by subtracting -0.044991895 mas from the nominal value according to \cite{Lindegren2021}. We used an upper limit of $A_{V} = 8.1537$ for TOI-6002 and $A_{V} = 0.0477$ for TOI-5713 on the extinctions from the dust maps of \cite{Schlafly&Finkbeiner2011} and a Gaussian prior on the stellar metallicity from  IRTF/SpeX (see Table \ref{table:stars}). Also, a Gaussian prior was applied on the 2MASS $m_{K}$ band to be used by \cite{Mann_2015} and \cite{Mann_2019} relations. We run the fit with $R_*$, $M_*$, and the effective temperature $T_{eff}$ as free parameters. The SED fits are presented in Fig.~\ref{fig:sed}, and results are reported in Table \ref{table:stars}.

\subsubsection{Galactic kinematics}
\label{sec:kinematics}

We used the \gaia{} DR3 data for each target, including coordinates, proper motions, parallaxes, and radial velocities \citep{GaiaCollaboration_2020}, to infer their $UVW$ galactic velocities in the local standard of rest (LSR) frame.
We used \texttt{astropy.coordinates} \citep{AstropyCollaboration2013, AstropyCollaboration2018, AstropyCollaboration2022} to convert the \gaia{} data to $UVW$ velocities, utilizing the correction of \citet{Schonrich2010} for the barycentric velocity relative to the LSR.
The resulting velocities are summarized in \autoref{table:spec+astro}.
Following Equations 6 and 7 of \citet{LiChengdong2017}, we find these $UVW$ velocities correspond to thin-disk, thick-disk, and halo membership probabilities of $\{63.9 \pm 1.3\%, 36.1 \pm 1.3\%, < 0.1\%\}$ and $\{97.2 \pm 0.1\%, 2.9 \pm 0.1\%, < 0.1\%\}$ for TOI-6002 and TOI-5713, respectively.
The likely thin-disk membership of both stars implies ages of $\lesssim 8.2$\,Gyr \citep{Kilic2017}.

\begingroup
\begin{table*} 
	\caption{
		Stellar astrometric and photometric properties of TOI-6002 and TOI-5713.} 
	\begin{center}
		\renewcommand{\arraystretch}{1.15}
		\begin{tabular*}{\linewidth}{@{\extracolsep{\fill}}l c c c}
			\toprule
			Parameter & TOI-6002 & TOI-5713 & Source                           \\
			\midrule
			TIC             & 102734241             & 219041246            & 1 \\
			2MASS           & J19582747+3217197     & J13534988+5236102    & 2 \\
			UCAC 4          & 612-092217            & 714-051779           & 3 \\
			\gaia\ DR3      & 2034047656176989952   & 1560676770454324224  & 4 \\
			\midrule
			\multicolumn{3}{c}{Photometry} \\   \\                      
			$TESS$	        & 12.6396 $\pm$ 0.0074  & 12.7022 $\pm$ 0.0073  & 1 \\
            $V$             & 14.6    $\pm$ 0.2     & 15.355 $\pm$ 0.018   & 3 \\
            $B$             & ...                   & 17.035 $\pm$ 0.011   & 3 \\
			$BP$            & 15.510  $\pm$ 0.030   & 15.58  $\pm$ 0.02    & 4 \\
			$Gaia$	        & 13.930  $\pm$ 0.030   & 13.66  $\pm$ 0.02    & 4 \\
			$RP$            & 12.710  $\pm$ 0.030   & 12.76  $\pm$ 0.02    & 4 \\
			$J$	            & 10.997  $\pm$ 0.023   & 11.069 $\pm$ 0.02    & 2 \\
			$H$	            & 10.425  $\pm$ 0.022   & 10.467 $\pm$ 0.02    & 2 \\
			$K$	            & 10.175  $\pm$ 0.019   & 10.217 $\pm$ 0.02    & 2 \\
			WISE 3.4 $\mu$m	& 9.940   $\pm$ 0.030   & 10.041 $\pm$ 0.03    & 5 \\
			WISE 4.6 $\mu$m	& 9.786   $\pm$ 0.030   & 9.877  $\pm$ 0.030   & 5 \\
			WISE 12  $\mu$m	& 9.459   $\pm$ 0.30    & 9.723  $\pm$ 0.038   & 5 \\
			WISE 22  $\mu$m	& 8.650   $\pm$ 0.407   & 9.019  $\pm$ 0.378   & 5 \\
            $yPS$           & ...                   & 12.307 $\pm$ 0.030   & 6 \\
            $zPS$           & ...                   & 12.634 $\pm$ 0.030   & 6 \\
			\midrule
			\multicolumn{3}{c}{Astrometry} \\ \\
			RA  (J2000)     & 19 58 27.46           & 13 53 50.10          & 4 \\
			DEC (J2000)     & +32 17 19.18          & +52 36 10.24         & 4 \\
			RA PM (mas/yr)  & -37.926 $\pm$ 0.015   & 108.546 $\pm$ 0.015  & 4 \\
			DEC PM (mas/yr) & -346.509 $\pm$ 0.016  & -19.281 $\pm$ 0.017  & 4 \\
                Radial velocity (km/s) & -10.20 $\pm$ 2.21 & -14.69 $\pm$ 1.64 & 4 \\
			Parallax (mas)  & 31.1664  $\pm$ 0.0149  & 24.3773 $\pm$ 0.0162 & 4 \\
                $U$ galactic velocity (km/s) & 3.78 $\pm$ 0.91 & 14.43 $\pm$ 0.88 & 6 \\
                $V$ galactic velocity (km/s) & -36.4 $\pm$ 1.6 & -32.3 $\pm$ 0.48 & 6 \\
                $W$ galactic velocity (km/s) & -42.9 $\pm$ 1.2 & -6.9 $\pm$ 1.3 & 6 \\
			\bottomrule
		\end{tabular*}
	\end{center}
 1. \citet{Stassun_2018}, 
 2. \citet{Cutri_2003}, 
 3. \citet{Zacharias_2012}, 
 4. \citet{GaiaCollaboration_2020}, 
 5. \citet{Cutri_2013}.
 6. This work.
	\label{table:spec+astro}
	
\end{table*}
\endgroup

\begingroup
\begin{table*}
\caption{Stellar parameters. Parameters in bold are the adopted stellar values in the analyses.}
\begin{center}
\renewcommand{\arraystretch}{1.15}
\begin{tabular*}{\linewidth}{@{\extracolsep{\fill}}l c c c @{}}
\toprule
Parameter                   & TOI-6002                    & TOI-5713                  & Source                     \\
\midrule
Sp. Type                    & M3.5$\pm$0.5                & M3.5$\pm$0.5              & IRTF/SpeX$^a$              \\
                            & M4$\pm$0.5                  & M4$\pm$0.5                & Shane/Kast$^a$             \\
\midrule
$T_{\rm eff}/{\rm K}$       & \textbf{3241$_{-60}^{+80}$}          & \textbf{3228$\pm$41}               & Joint fit$^b$              \\
		                    & 3477$\pm$100                & 3277$\pm$100              & \cite{Mann_2015}           \\
		                    & 3314$\pm$80                 & 3300$\pm$80               & Stefan-Boltzmann           \\
\midrule
$\mathrm{[Fe/H]}$           & \textbf{-0.18 $\pm$ 0.12}   & \textbf{-0.03 $\pm$ 0.13} & IRTF/SpeX$^c$             \\
                            & {+0.04 $\pm$ 0.20}          &  {+0.08 $\pm$ 0.20}       & Shane/Kast$^d$            \\
\midrule
$M_\star/\msun$             & \textbf{0.2409$_{-0.0065}^{+0.0066}$}& \textbf{0.2654$\pm$0.0061}         &  Joint fit$^b$             \\
\midrule
$R_\star/\rsun$             & \textbf{0.2105$_{-0.0048}^{+0.0049}$}& \textbf{0.2985$_{-0.0073}^{+0.0074}$} & Joint fit$^b$           \\
\midrule
$L_\star.10^{-3}/\lsun$     & \textbf{5.80$_{-0.38}^{+0.51}$}      & \textbf{8.73$_{-0.29}^{+0.27}$}    &  Joint fit$^b$             \\
                            & 6.39126$\pm$0.5621          & 9.1564$\pm$0.8340         & $BC_{K}$ \citep{Mann_2015} \\            
\midrule
$\log g_\star / {\rm dex}$  & \textbf{4.998$\pm$0.025}             & \textbf{4.912$\pm$0.023}           &  Joint fit$^b$             \\
\midrule
$\rho_\star$ / g\,cm$^{-3}$ & \textbf{21.2$_{-1.7}^{+1.9}$}        & \textbf{14.1$_{-1.0}^{+1.1}$}      &  Joint fit$^b$             \\
                            & $23.4_{-3.2}^{+1.4}$        & $11.4_{-6}^{+7}$          & Transits fit alone         \\
\midrule
Age / Gyr                   & $\gtrsim 4.5$                & $\lesssim 4.5$             & H$\alpha$ emission         \\
                            & $\lesssim 8.2$               & $\lesssim 8.2$             & Kinematics \\
                            & $> 2.19 \pm 0.10$           & $< 1.81 \pm 0.10$           & Mass--spindown relation \citep{Pass2024} \\
\bottomrule
\end{tabular*}
\end{center}

$^a$ Classification based on spectral templates from \citet{2007AJ....133..531B} and spectral type/index relations from \citet{1995AJ....110.1838R}, \citet{1999AJ....118.2466M}, \cite{2003AJ....125.1598L} and \citet{2007MNRAS.381.1067R}. \\
$^b$ A joint fit of the spectral energy distribution (SED) and transits with the use of \cite{Mann_2015} and \cite{Mann_2019} empirical relations (see Sect.~\ref{JointSED} and \ref{Modeling}). \\
$^c$ Metallicity based on measurement of $K$-band Na\,\textsc{i} and Ca\,\textsc{i} doublets and the H2O--K2 index \citep{Rojas-Ayala2012}, and the calibrations of \citet{Mann2014}. \\
$^{d}$ Metallicity based on measurement of the $\zeta$ index \citet{2007ApJ...669.1235L,2013AJ....145...52M} and calibrations from \citet{2013AJ....145...52M}. \\
 \label{table:stars}
\end{table*}
\endgroup

\subsection{Discarding possible false-positive scenarios}

\begin{figure}[h!] 
	\includegraphics[width=\columnwidth]{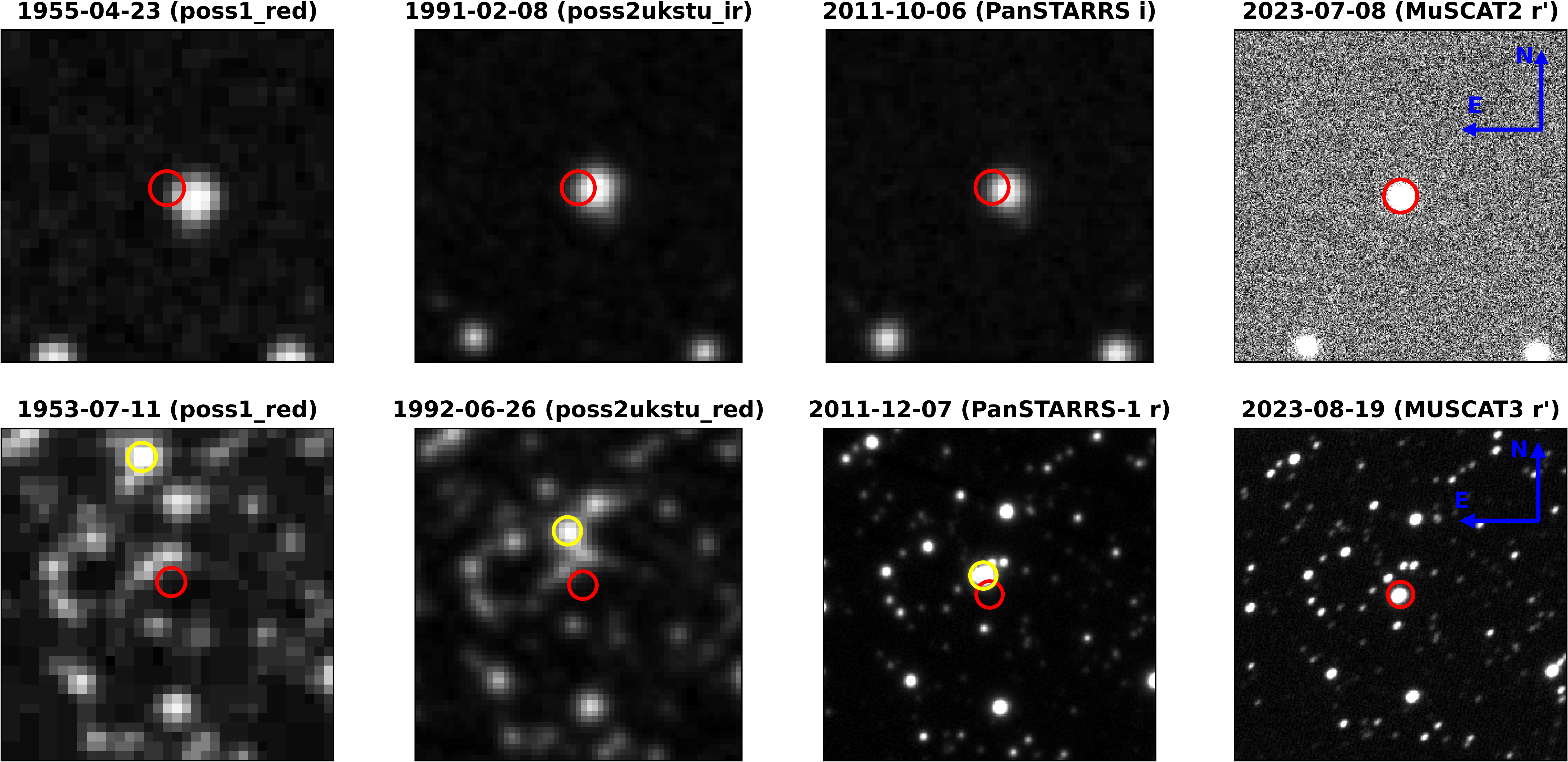}
	\caption{Archival images cropped with a field of view of 1’×1’ around TOI-5713
(top row) and TOI-6002 (bottom row). Circles in red show the current position of the stars, and the circles in yellow, in the bottom row, show the positions of TOI-6002 at the dates of exposures.}
	\label{fig:image-arx}
\end{figure}

Two primary cases of astrophysical false positives can masquerade as exoplanet transits in TESS data. The first case is background/foreground eclipsing binaries (EBs), especially those in grazing configurations. For both targets, we searched for archival imaging, retrieving images from POSS I/DSS \citep{1963POSS-I} and POSS II/DSS \cite{1996DSS_POSS-II}. These images, in addition to those taken from our recent observations from MusCAT3 (for TOI-6002) and MusCAT2 (for TOI-5713) in 2023, span more than 70 years, as shown in Fig.~\ref{fig:image-arx}. TOI-6002 has a high proper motion of $\sim$344.42~mas/yr and has moved by $\sim$24.4\arcsec\ from 1955 to 2023. This allowed us to confirm that there was no bright star in the current position of TOI-6002. On the other hand, TOI-5713 has a relatively low proper motion ($\sim$110.24~mas/yr) and has moved by $\sim$7.7\arcsec\ from 1953 to 2023, making it challenging to confirm the absence of any unresolved companion in its current position. However, high-resolution adaptive optics imaging from Keck-II (for TOI-5713) and Palomar (for TOI-6002) confirm that no stellar companions were detected close to ($\lesssim 1\arcsec$) the targets as discussed in Sect.~\ref{OBS:AO}. Additionally, the renormalized unit weight error (RUWE) from \gaia\ DR3 is below 1.4 for both stars, indicating that the stars are singles (see Sect.~\ref{sub:gaia}).

The second astrophysical false positive case is that of bound eclipsing binaries. Many findings argue against this scenario for both stars: (1) the spectral energy distribution is fit well by single-star models (see Fig.~\ref{fig:sed}); (2) the spectra collected by SpeX SXD and Shane/Kast, for both stars, show no secondary spectrum (see Sect.~\ref{table:spec+astro}), and Fig.~\ref{fig:spex} and \ref{fig:kast}; (3) the achromatic nature of the transits, where their depths in all bands agree within $1\sigma$ for each planet (see Table~\ref{fit:results_combined}); (4) the stellar densities determined from the transits alone are in good agreement with the those determined from the the basic parameters of each star (see Table \ref{table:stars}).
Considering these arguments, we discard these false-positive scenarios.

\subsection{Statistical validation}

In addition to ruling out false-positives via the arguments in the previous section, we used the Tool for Rating Interesting Candidate Exoplanets and Reliability Analysis of Transits Originating from Proximate Stars \citep[\texttt{TRICERATOPS:}][]{Giacalone_2021} to statistically validate TOI-6002\,b and TOI-5713\,b. 
TRICERATOPS simulates potential astrophysical false positives due to bound stellar companions, nearby stars, and line-of-sight stars.
To assess the planetary nature of the transit signal, it calculates two quantities:
the false positive probability (FPP), which quantifies the probability that the signal originates from other astrophysical sources; 
and the nearby false positive probability (NFPP), which quantifies the probability that the transit signal originates from a resolved nearby star rather than the target star. 
The thresholds for statistically validating a planet candidate are $\mathrm{FPP} < 0.01$ and $\mathrm{NFPP} < 0.001$. 

We ran \texttt{TRICERATOPS} on TESS data supplied with contrast curves from NIRC2 speckle imaging for TOI-5713 and from Palomar for TOI-6002. As our ground-based photometry confirmed the transit events were on target, removing the possibility of a nearby false positive, we set the NFPP to 0. The inferred FPP is 0.0024$\pm$0.001 for TOI-5713\,b 
and 0.00783$\pm$0.00016 for TOI-6002\,b. Independently, we also ran \TRICERATOPS\ on the data obtained with MuSCAT3 (for TOI-6002\,b) and MuSCAT2 (for TOI-5713\,b) because they present a  higher photometric precision. Also supplied with contrast curves, this led to FPP and NFPP lower than $6\times10^{-4}$ for both candidates. Given these results using both TESS and  ground-based data independently, we consider the two candidates as statistically validated 
planets.

\subsection{Planet search and detection limits}

As mentioned in Sect.~\ref{TESS:phot}, TOI-5713 and TOI-6002 have both been observed in multiple sectors of TESS --- respectively, sectors 16, 22, 23, 46, and sectors 14, 41, 54, 55. Since both planets were detected by the TESS pipelines with a low signal-to-noise ratio, it is possible additional planets remained unnoticed below the threshold set to trigger an alert (Multiple Event Statistic = 7.1$\sigma$) \citep{Tenenbaum_2012}. 
In this context, we performed a search for additional planet candidates in the TESS data with the \texttt{SHERLOCK}\footnote{\url{https://github.com/franpoz/SHERLOCK}} pipeline \citep{2020_sherlock,2020_sherlock_demory,sherlock2024} and assessed the detection limits with injection and recovery tests with the \texttt{MATRIX}\footnote{\url{https://github.com/PlanetHunters/tkmatrix}} ToolKit \citep{2020_sherlock,2022_matrix}. 

The \texttt{SHERLOCK} pipeline is an open-source package that provides six different modules to find and validate transit signals in photometric time-series observations: (1) acquiring automatically the data from an online database, such as MAST in the case of TESS data; (2) searching for planetary transit signals; (3) performing a vetting of the detected signals; (4) conducting a statistical validation of the vetted signals; (5) modeling them to retrieve their ephemerides; and (6) finding the upcoming transit windows observable from ground-based observatories. See \cite{Delrez2022} and \cite{pozuelos2023} for recent applications and further details.  

\texttt{SHERLOCK} obtains the PDC-SAP fluxes from the MAST archive, and applies a bi-weight filter using the \texttt{W{\={o}}tan} package \citep{2019_wotan_hippke} to detrend the light curve with various window sizes. The transit search is then performed over the nominal light curve as well as the detrended ones using the \texttt{Transit Least Square} (TLS) package \citep{2019_TLS_Hippke_Heller}. This strategy maximizes the signal-to-noise ratio (S/N) and the signal detection efficiency (SDE) of the transit signals found. In the case of TOI-6002 and TOI-5713, we downloaded the short-cadence data from TESS and applied seven different window sizes between 0.2 and 1.0 days to detrend the data. As the transit search is performed in an iterative way, if a signal is detected with a S/N$\geqslant$5, it is masked, and a new search is implemented. The operation is repeated until the algorithm no longer finds signals above the threshold or reaches the number of loops (so-called ``runs'') set by the user. For all our analyses, we kept the number of detrends to 7, set the maximum number of runs to 5, and also kept the S/N threshold to 5. 

We performed a first test to see if the first planet is recovered well by \texttt{SHERLOCK} for both systems with a period grid between 1 and 15 days. For TOI-6002, the signal was not recovered in the first instance. We then applied a Savitzky-Golay digital filter \citep{1964_SG_filter} and repeated the test, where we indeed recovered the planet. TOI-6002 b was recovered in the second run and in seven of the detrended light curves with a maximum S/N of 12.8 and an SDE of 13.0. Similarly, TOI-5713 b was recovered in the first run and in six of the detrended light curves and, at best, with an S/N of 11.8 and an SDE of 23.1. In order to assess the presence of additional candidates, we performed a search for the period grid between 0.5 and 25 days on 7 light curves detrended by a bi-weight filter with window sizes from 0.2 and 1.2 days, allowing a maximum of 5 runs. Neither of these analyses yielded positive planetary-like signals; all signals could be attributed to noise, variability, or systematics. 

The lack of additional transiting planet candidates in the TESS data for both systems does not necessarily exclude the existence of extra planets. Several scenarios could explain the lack of interesting signals in our transit search with \texttt{SHERLOCK}: (1) the systems only host one planet; (2) if other planets exist, they are not transiting; (3) the data we examined do not have the photometric precision required to detect the additional transiting planets, or they have a longer period than the period grid searched in our analyses \citep{Wells2021}. In the case of scenario (2), radial velocity measurements may be able to detect them, as discussed in Sect.~\ref{RV_}. In the case of scenario (3), a ground-based high-precision photometric follow-up could possibly detect more planets, such as was the case for LP 890-9 c \citep{Delrez2022}. However, to test scenario (3) against the TESS data, we performed injection and recovery tests over the available short-cadence data for both systems.

\texttt{MATRIX} is an injection and recovery software specifically built to establish the retrieval thresholds of hypothetical transiting exoplanet signals in space-based data for a given star. It is built on top of the same core features as \texttt{SHERLOCK} so that it can reproduce the same algorithms and techniques implemented in the pipeline, and their results can be compared. That is, \texttt{MATRIX} sets detection limits to \texttt{SHERLOCK}, and by extension, to any similar pipeline using preliminary detrending and phase-folded least squares search techniques. To do so, three-dimensional grids (orbital period, planet size, orbital phase) are built to create synthetic scenarios. For each of these scenarios, \texttt{MATRIX} runs the pre-processing of the curve (smooth, noise-masks, detrending) and searches for the injected planet either using TLS or Box Least Squares \citep[BLS;][]{2002_bls}.

\begin{figure}[h!] 
	\includegraphics[width=\columnwidth]{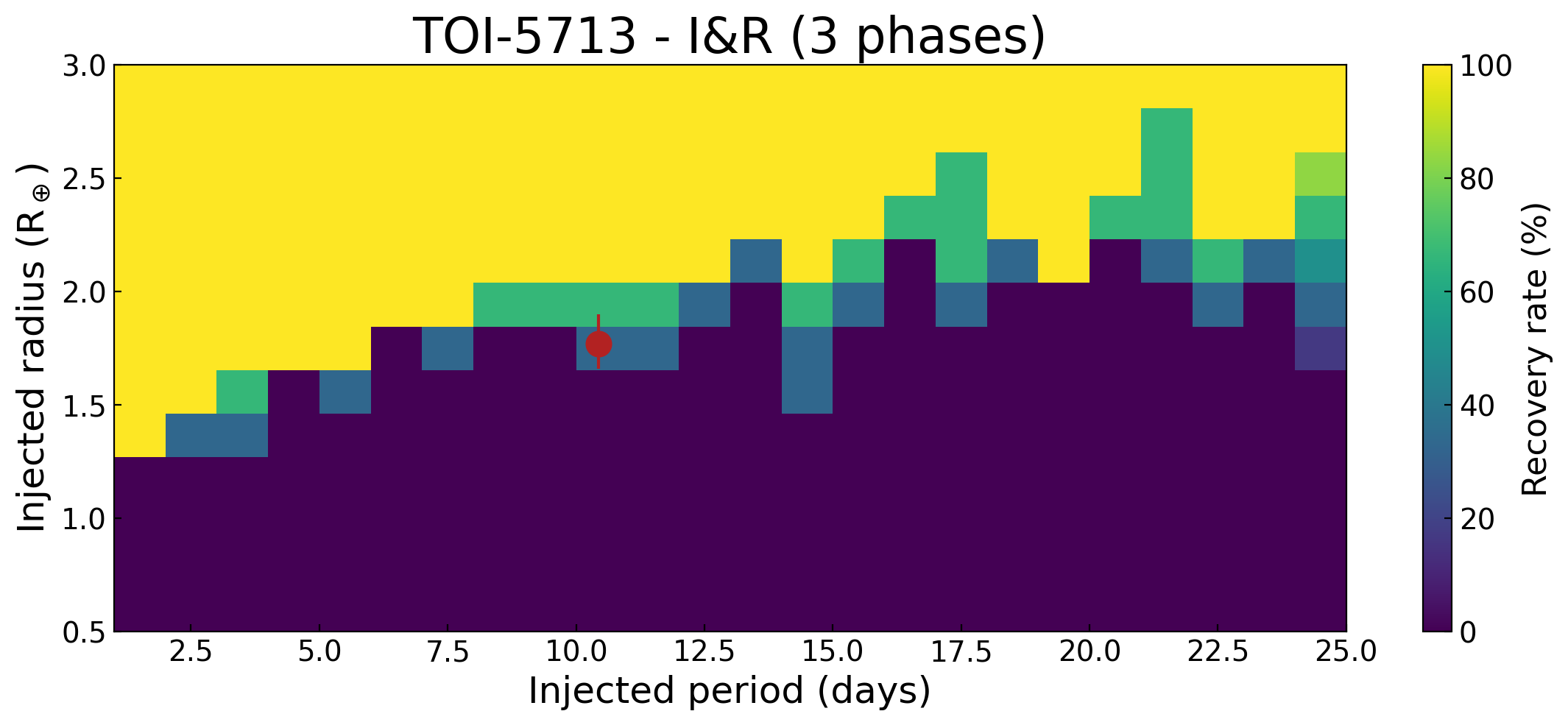}
	\includegraphics[width=\columnwidth]{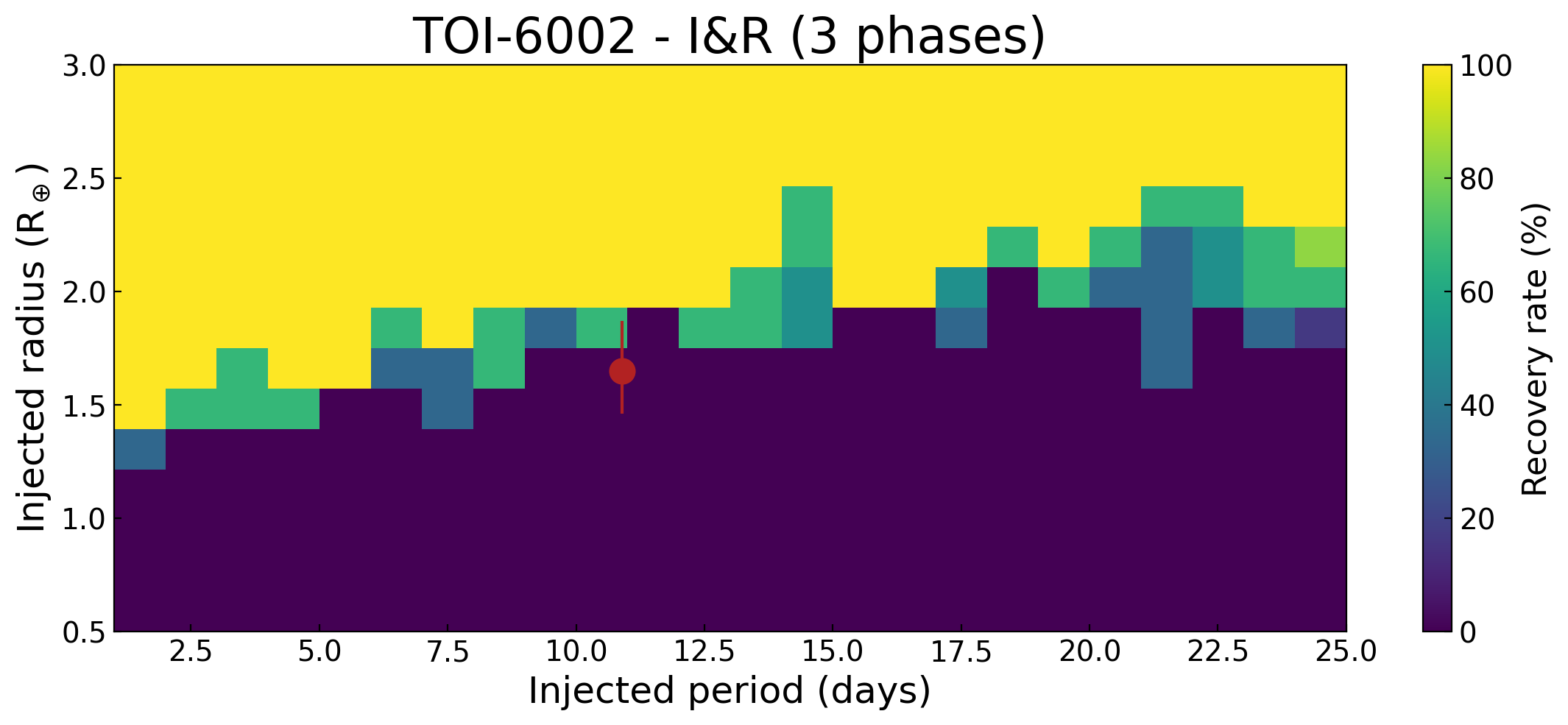}
	\caption{Top panel: TOI-5713 injection and recovery test, where TOI-5713 b is rendered with a red marker. Bottom panel: TOI-6002 injection and recovery test, where TOI-6002 b is rendered with a red marker.}
	\label{fig:inj-rec}
\end{figure}

As both systems' planets are very similar in orbital periods and sizes, we planned the same scenarios for both of them. We wanted to explore the recoverability of planets with sizes between 0.5 and 3 $R_\oplus$ and orbital periods from 1 up to 25 days. Therefore, the final injection and recovery test cases included a grid of 25 periods (from 1 to 25 days), 15 planet radii (from 0.5 to 3 $R_\oplus$), and 3 orbital phases, summing a total of 1125 independent injected synthetic transiting planets for each target. 

Our tests show that TOI-5713 b and TOI-6002 b lie at the detectability limits found for each of their systems (see Fig.~\ref{fig:inj-rec}). This is consistent with the low S/N obtained for both planets, either in \texttt{SHERLOCK} ($S/N_\mathrm{TOI-5713 b}=11.8$, $S/N_\mathrm{TOI-6002 b}=12.8$) or \texttt{SPOC} ($S/N_\mathrm{TOI-5713 b}=7.7$, $S/N_\mathrm{TOI-6002 b}=8.1$), and the low number of \texttt{SHERLOCK} detrends where they were found. It is essential to recognize that the S/N reported by the SHERLOCK algorithm is derived from the TLS package, which employs a straightforward estimation approach that is not directly comparable to the metrics produced by alternative pipelines, such as SPOC. Instead, SHERLOCK utilizes these values internally to evaluate the detected signals and identify the most significant ones among them.

From these results, we rule out the existence of transiting planets larger than $2.3\,R_\oplus$ with orbital periods up to 25 days around TOI-6002, whilst the same statement applies for planets larger than $3\,R_\oplus$ in TOI-5713. If additional transiting exoplanets were present in the analyzed period range, they would probably be within the limits or below the super-Earth radius regime in TOI-6002. On the other side, there could also be undetected sub-Neptune transiting exoplanets in TOI-5713, especially for outermost orbits above 14 days, where planets with a radius between $2\,R_{\oplus}$ to $2.75\,R{_\oplus}$ could remain undetected with detection rates between 33--66\%.

\subsection{Stellar activity and rotation period}\label{sec:activity}

If the host star is magnetically active and has an inhomogeneous photosphere, its rotation can pose a challenge for accurately measuring an exoplanet's mass via radial velocity (RV) observations, especially if the rotation time scale is similar to the orbital period of the planet. Therefore, we examined the effect of stellar activity on RV observations. 

We first visually inspected the PDC-SAP data light curves of the two targets. We observed no rotational modulations nor flares for TOI-6002, while we found clear flares in the four observed sectors of TOI-5713 in addition to hints of rotational modulation. 

\begin{figure}[h!] 
	\includegraphics[width=\columnwidth]{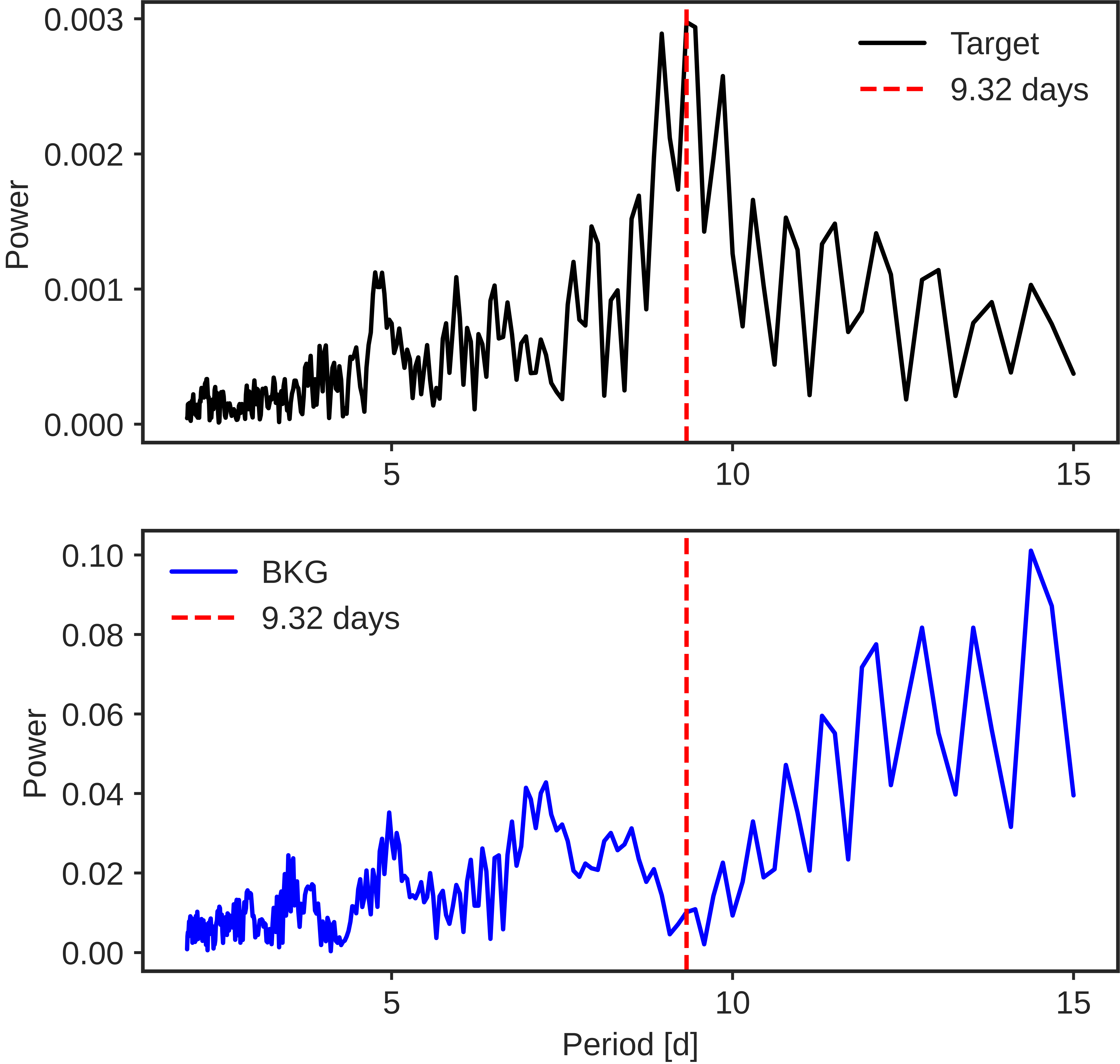}
	\caption{\texttt{TESS-SIP} power spectrum of TOI-5713 based on the four TESS sectors. We calculate the periodogram for both the corrected light curve (top panel) and the background (BKG) pixels (bottom panel). The star's periodogram shows some power excess compared to the BKG.}
	\label{fig:TESS-SIP}
\end{figure}

We then used the Systematics-Insensitive Periodogram
(\texttt{TESS-SIP}\footnote{\url{https://github.com/christinahedges/TESS-SIP}}) \citep{Hedges2020} to build a periodogram using the whole TESS photometry (4 sectors) for each of the two targets. This tool simultaneously creates a Lomb-Scargle periodogram and detrends TESS systematics as described in \cite{Angus2016}. The SPOC pipeline is known to attenuate rotational periods, thus \texttt{TESS-SIP} reproduces simple aperture photometry (SAP) light curves from the TPFs and the SPOC apertures. We applied this procedure on all four sectors observed by TESS. Searching for periods between 2 and 15 days, we applied \texttt{TESS-SIP} on the target and on all the background pixels around the SPOC apertures. We found no significant rotational signal for TOI-6002; however, we found a significant rotational period at 9.32$\pm$0.20 days for TOI-5713. Thus, we report the SIP power spectrum for only TOI-5713 in Fig.~\ref{fig:TESS-SIP}. We also found that this signal remains on the PDC-SAP light curves of each individual sector, which underscores the robustness of the signal. Beyond 15 days, we did not find any rotational signal.  

Given these rotation-period constraints, we can estimate the ages of the stars using the mass-dependent spindown relation from \citet{Pass2024}.
The ``jump'' times at which fully convective M dwarfs transition from rapidly to slowly rotating are $2.19 \pm 0.10$\,Gyr and $1.81 \pm 0.10$\,Gyr for stars with the masses of TOI-6002 and TOI-5713, respectively \citep[see][Equation 2]{Pass2024}.
Assuming that our non-detection of rotational modulations for TOI-6002 owes to a long rotational period ($> 10$\,d), we can constrain its age to $> 2.19 \pm 10$\,Gyr.
By contrast, the relatively rapid rotation of TOI-5713 implies an age of $< 1.813 \pm 0.10$\,Gyr.
These estimates are consistent with the age estimates we derive from their H$\alpha$ emission (for TOI-5713) or lack of it (for TOI-6002) as well as the likely thin-disk membership of both stars.

\subsection{Global modeling}
\label{Modeling}

For each planet, we used the latest version of \texttt{EXOFASTv2} \citep{Eastman_2019} software package to globally analyze the TESS and ground-based data described in Sect.~\ref{obs:phot}, jointly with the star's SED fit and the empirical relations \cite{Mann_2015} and \cite{Mann_2019} to characterize the host stars (see description in Sect.~\ref{JointSED}).

\begin{figure*} 
\centering
\includegraphics[width=\textwidth]{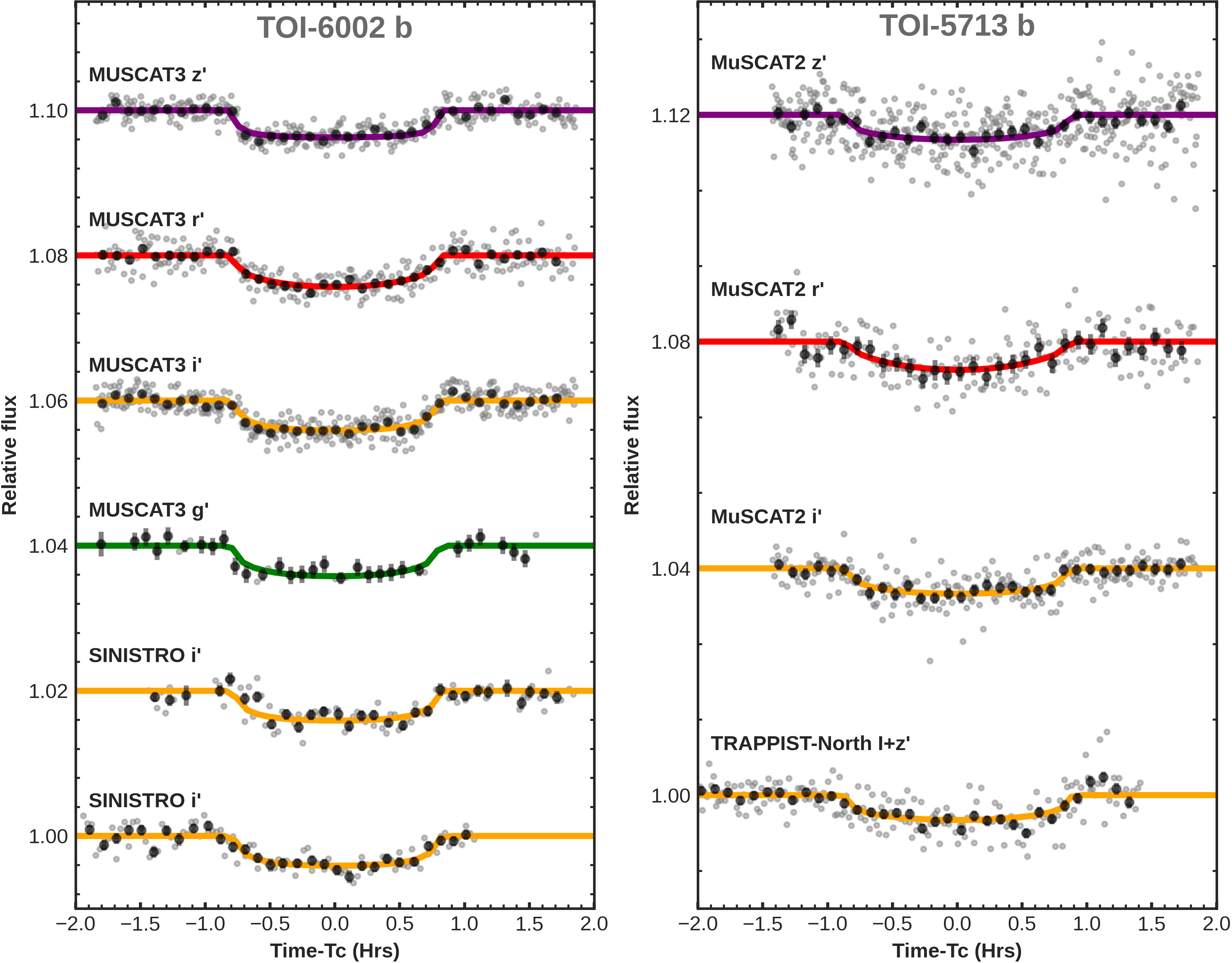}
\caption{Phase-folded detrended ground-based transit light curves with time in hours from mid-transit of TOI-6002 b (left) and TOI-5713 b (right). The unbinned light curves are shown with gray points. The binned points
(bin size=6 min) are shown in black with the corresponding error bars. The solid-colored lines corresponds to the best-fit transit model from the final joint fit.}
\label{GB:LC}
\end{figure*}

\begin{figure*} 
	\includegraphics[scale=0.62]{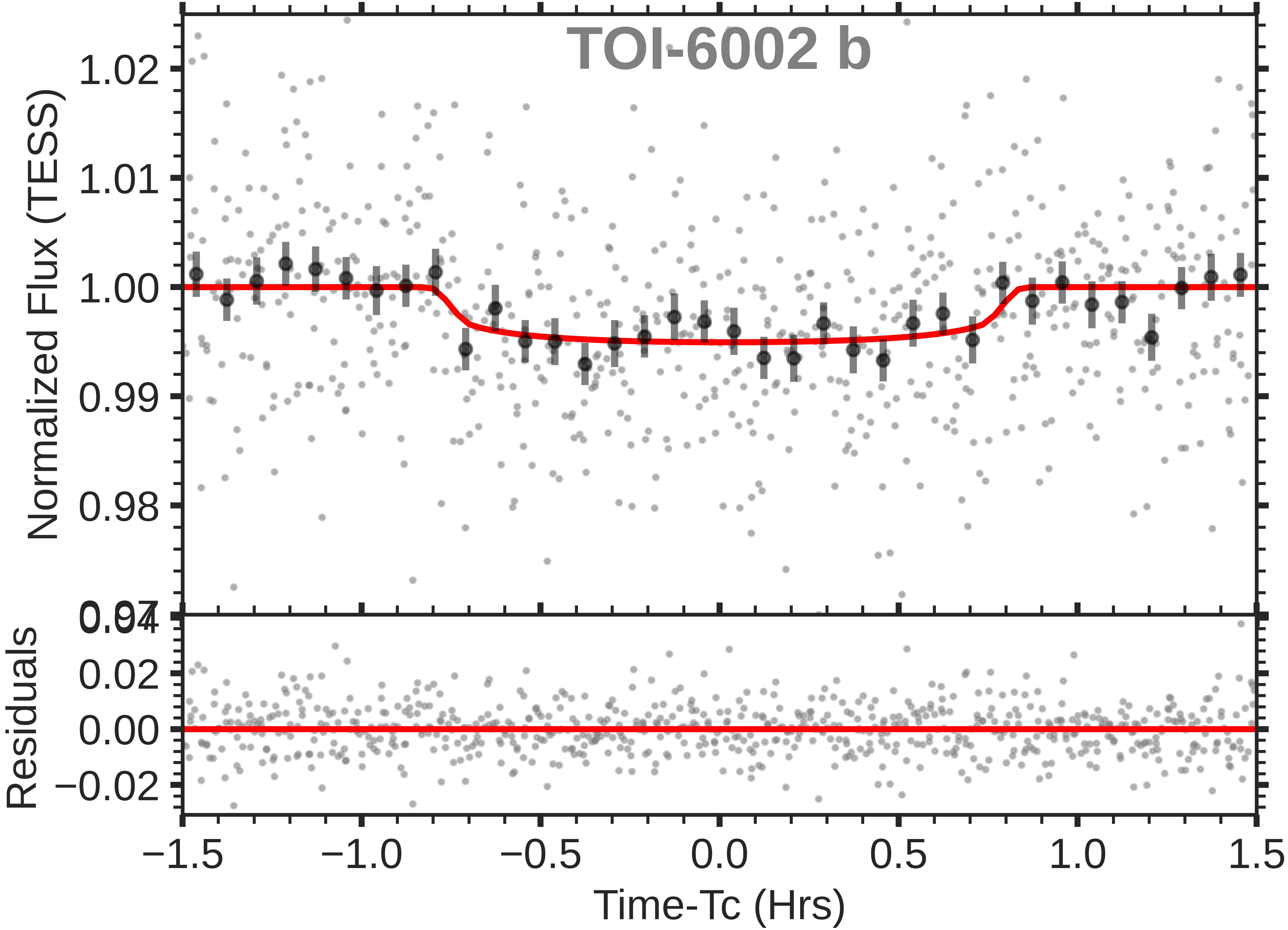}
	\includegraphics[scale=0.62]{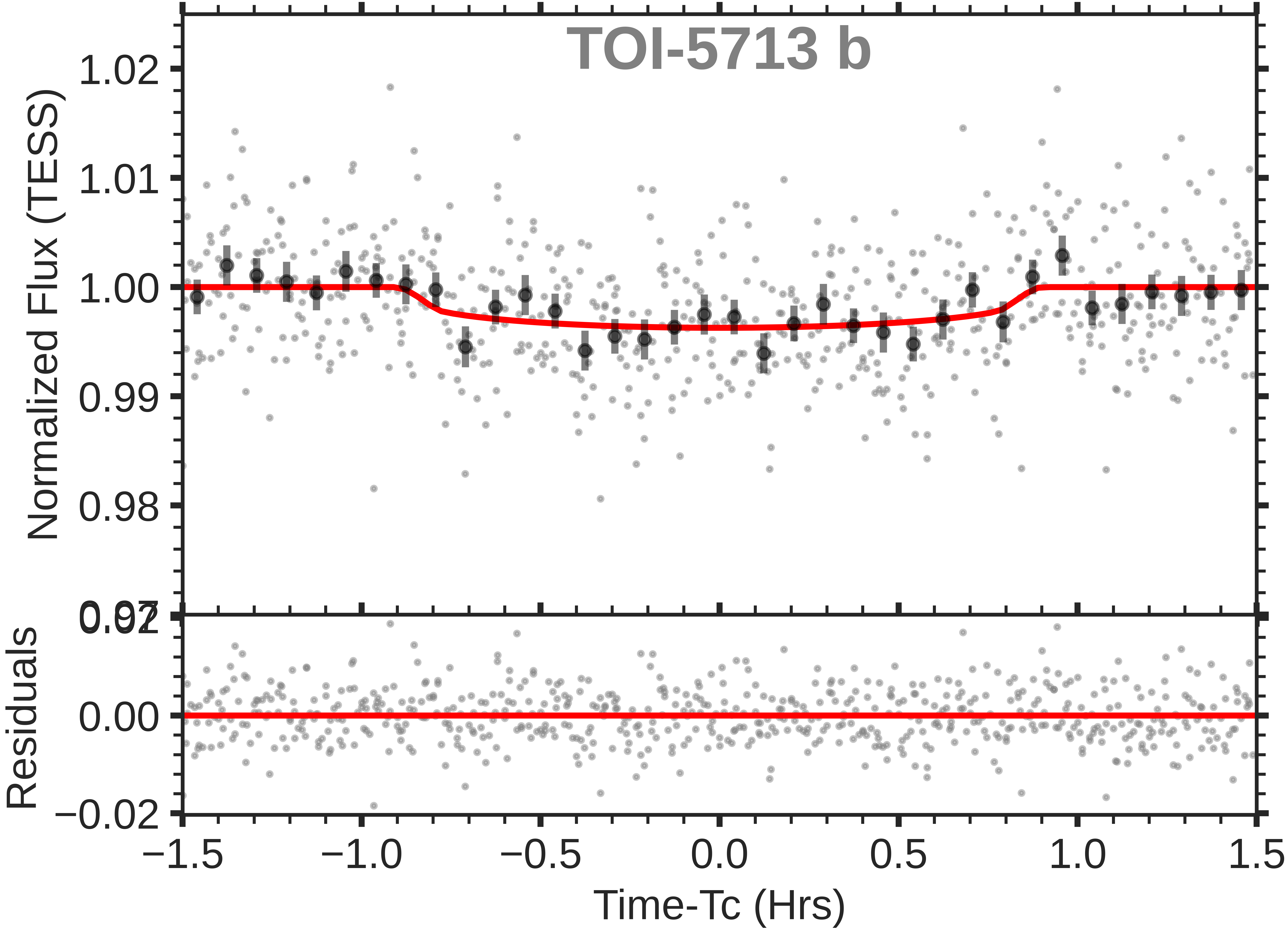}
	\caption{TESS phase-folded detrended light curve with the time in hours from mid-transit of TOI-6002 b (left) and TOI-5713 b (right). The red solid lines represent the best-fit transit model from the global fit. The gray dots are the 2-min TESS data and the black dots with error bars show the data binned every 5 minutes. The residuals are plotted in the bottom panel.}
	\label{fig:TESS_PH}
\end{figure*}

In the priors file, we set starting values of $M_*$, $R_*$ and $T_{eff}$. We set lower and upper limits priors on the orbital period ($P \pm 0.1P$ d) and the transit epoch ($T_c\pm 0.33P$) from the values determined by SPOC and reported in ExoFOP. For the LD parameters, we disabled the priors based on the Claret tables \citep{Claret_2017,Claret&Bloemen_2011} by setting the \texttt{NOCLARET} flag given the low mass of our stars where these tables are not reliable \citep[see][and references therin]{Eastman_2019}. Instead, we fitted for quadratic LD parameters assuming Gaussian priors of variance 0.1, computed with the \texttt{PyLDTk} code \citep{Parviainen_2015MNRAS} for each passband (see Table \ref{tab:ld}). Regarding the TESS data, we used the detrended PDC-SAP light curves (see Sect.~\ref{TESS:phot}) which are corrected for the contribution of neighboring stars and subsequently corrected for sky background bias for TOI-6002 in sector 14 (see Sect.~\ref{TESS:phot}), but we still fit for a possible dilution by applying a Gaussian prior of 0$\pm$10\% of the contamination ratio reported in ExoFOP. This is to propagate any uncertainty in the correction. Also, Gaussian priors were applied on the metallicities from IRTF/SpeX (see Table~\ref{table:stars}). For each planet, we fitted for circular and eccentric orbits to check for orbital eccentricity evidences using only photometric data. The two fits were run until convergence. We found the difference in the loglikelihood is $\Delta\ln Z = \ln Z_\mathrm{eccentric} - \ln Z_\mathrm{circular} \geqslant 2$ for both TOI-6002\,b and TOI-5713\,b, which suggests that eccentric orbits are marginally favored for both planets. Table \ref{fit:results_combined} presents the fits' results for both the circular and eccentric cases. Fig.~\ref{GB:LC} shows the detrended and modeled gound-based light curves for both TOI-6002 b and TOI-5713 b, and Fig.~\ref{fig:TESS_PH} shows the TESS phase-folded light curves.  

We also re-fitted the data for a circular orbit and with the same priors as explained above but without any Gaussian prior on the stellar mass. This is to check for consistency between the stellar densities determined from $R_*$ and $M_*$ (from SED + empirical relations) and those determined from the transits via the $a/R_*$ (the semi-major axis over the stellar radius) ratio. We found  $\rho_* = 23.4^{+1.4}_{-3.2}$\,g\,cm$^{-3}$   for TOI-6002\,b and $11.4_{-6}^{+7}$\,g\,cm$^{-3}$ for TOI-5713\,b, which are consistent with the previously determined stellar densities (see Table \ref{table:stars}). This provides additional assurance that the transit signals for both candidates originate from transiting planets and not astrophysical false positives. 

\begin{table}
\caption{Priors assumed for the quadratic limb-darkening coefficients for each passband for both stars TOI-6002 and TOI-5713.}\label{tab:ld}
\begin{center}
\begin{tabular}{l c c l}
\toprule
Filter &  $u_1$ & $u_2$ \\
\midrule
TOI-6002 \\
\midrule
TESS   & 0.2534$\pm$0.1 & 0.3798$\pm$0.1 \\
$g'$   & 0.4895$\pm$0.1 & 0.4039$\pm$0.1 \\
$r'$   & 0.4992$\pm$0.1 & 0.3384$\pm$0.1 \\
$i'$   & 0.2856$\pm$0.1 & 0.4168$\pm$0.1 \\
$z_s$   & 0.1907$\pm$0.1 & 0.4718$\pm$0.1 \\
\midrule
TOI-5713 \\
\midrule
TESS   & 0.2806$\pm$0.1 & 0.3069$\pm$0.1 \\
$r'$   & 0.5388$\pm$0.1 & 0.2842$\pm$0.1 \\
$i'$   & 0.3186$\pm$0.1 & 0.3422$\pm$0.1 \\
$z_s$   & 0.2383$\pm$0.1 & 0.3334$\pm$0.1 \\
$I+z$  & 0.2222$\pm$0.1 & 0.2880$\pm$0.1 \\
\hline 
\end{tabular}
\end{center}
\end{table}

{\renewcommand{\arraystretch}{1.4}
	\begin{table*}[!htbp]
		\caption{Median values and 68\% confidence intervals for the parameters of TOI-6002\,b and TOI-5713\,b obtained using EXOFASTv2 for circular and eccentric orbits.}
		\begin{center}
			{\small %
				\begin{tabular}{llcccc}
					\hline
					Parameter                 & Description                            & \multicolumn{2}{c}{TOI-6002\,b} & \multicolumn{2}{c}{TOI-5713\,b} \\ 
					\cline{3-6}
					& & Circular & Eccentric & Circular & Eccentric \\ 
					\hline
					\hline
					\smallskip\\\multicolumn{2}{l}{Stellar Parameters:}                        & \smallskip & & \smallskip & \\
					
					~~~~$M_*$\dotfill                 & Mass (\msun)\dotfill                   & $0.2105^{+0.0049}_{-0.0048}$ & $0.2101^{+0.0049}_{-0.0048}$ & $0.2653^{+0.0062}_{-0.0061}$ & $0.2653\pm0.0061$ \\
					~~~~$R_*$\dotfill                 & Radius (\rsun)\dotfill                 & $0.2409^{+0.0066}_{-0.0065}$ & $0.2432^{+0.0063}_{-0.0065}$ & $0.2984\pm0.0073$ & $0.2992^{+0.0073}_{-0.0072}$ \\
					~~~~$L_*$\dotfill                 & Luminosity ($10^{-3}$\lsun)\dotfill    & $5.80^{+0.51}_{-0.38}$ & $5.82^{+0.48}_{-0.36}$ & $8.74^{+0.27}_{-0.28}$ & $8.74^{+0.28}_{-0.29}$ \\
					~~~~$\rho_*$\dotfill              & Density (cgs)\dotfill                  & $21.2^{+1.9}_{-1.7}$ & $20.6^{+1.8}_{-1.5}$ & $14.1^{+1.1}_{-1.0}$ & $13.96^{+1.1}_{-0.99}$ \\
					~~~~$\log{g}$\dotfill             & Surface gravity (cgs)\dotfill          & $4.998\pm0.025$ & $4.989^{+0.025}_{-0.024}$ & $4.912\pm0.023$ & $4.910^{+0.023}_{-0.022}$ \\
					~~~~$T_{\rm eff}$\dotfill         & Effective Temperature (K)\dotfill      & $3241^{+82}_{-60}$ & $3229^{+77}_{-57}$ & $3229^{+41}_{-40}$ & $3225^{+41}_{-40}$ \\
					
					\smallskip\\\multicolumn{2}{l}{Planetary Parameters:}                  & \smallskip & & \smallskip & \\
					
					~~~~$P$\dotfill           & Orbital period (days)\dotfill                  & $10.904821^{+0.000021}_{-0.000018}$ & $10.904821^{+0.000021}_{-0.000018}$ & $10.441988^{+0.000014}_{-0.000013}$ & $10.441989^{+0.000015}_{-0.000014}$ \\
					~~~~$R_P$\dotfill         & Planet radius (\re)\dotfill                    & $1.63^{+0.21}_{-0.19}$ & $1.65^{+0.22}_{-0.19}$ & $1.75^{+0.16}_{-0.14}$ & $1.77^{+0.17}_{-0.14}$ \\
					~~~~$T_C$\dotfill         & Transit time     (\bjdtdb)\dotfill             & $2458692.7636^{+0.0024}_{-0.0028}$ & $2458692.7639^{+0.0024}_{-0.0028}$ & $2458745.6773^{+0.0014}_{-0.0016}$ & $2458745.6776^{+0.0015}_{-0.0017}$ \\
					~~~~$a$\dotfill           & Semi-major axis (AU)\dotfill                   & $0.05725\pm0.00044$ & $0.05722\pm0.00044$ & $0.06008\pm0.00046$ & $0.06008\pm0.00046$ \\
					~~~~$i$\dotfill           & Inclination (Degrees)\dotfill                  & $89.657^{+0.14}_{-0.099}$ & $89.68\pm0.21$ & $89.385^{+0.10}_{-0.088}$ & $89.37^{+0.36}_{-0.21}$ \\
					~~~~$e$\dotfill           & Eccentricity \dotfill                          & 0 (Fixed) & $0.15^{+0.42}_{-0.11}$ & 0 (Fixed) & $0.24^{+0.40}_{-0.17}$ \\
					~~~~$\tau_{circ}$         & Tidal circularization timescale (Gyr)          & $6200^{+4200}_{-2400}$ & $3500^{+3900}_{-3300}$ & $4600^{+2300}_{-1500}$ & $2000^{+2600}_{-2000}$ \\
					~~~~$\omega_*$            & Argument of Periastron (Degrees)               & 90 (fixed) & $-160^{+140}_{-110}$ & 90 (fixed) & $-139^{+120}_{-99}$ \\
					~~~~$T_{\rm eq}$\dotfill        & Equilibrium temperature$^{a}$ (K)\dotfill& $321.1^{+7.0}_{-5.6}$ & $321.4^{+6.6}_{-5.2}$ & $347.2^{+3.0}_{-3.1}$ & $347.2^{+3.0}_{-3.2}$ \\
					~~~~$R_P/R_\star$\dotfill       & Planet-to-star radius ratio\dotfill      & $0.0622^{+0.0080}_{-0.0070}$ & $0.0622^{+0.0080}_{-0.0070}$ & $0.0539^{+0.0048}_{-0.0039}$ & $0.0541^{+0.0050}_{-0.0041}$ \\
					~~~~$a/R_\star$\dotfill         & Semi-major axis in stellar radii\dotfill & $51.1^{+1.5}_{-1.4}$ & $50.6^{+1.4}_{-1.3}$ & $43.3\pm1.1$ & $43.2^{+1.1}_{-1.0}$ \\
					~~~~$\delta$\dotfill            & $\left(R_P/R_\star\right)^2$ (ppt)\dotfill& $3.87^{+1.1}_{-0.82}$ & $3.87^{+1.1}_{-0.83}$ & $2.90^{+0.54}_{-0.40}$ & $2.92^{+0.56}_{-0.42}$ \\
					~~~~$\delta_{\rm g'}$\dotfill   & Transit depth in $g'$ (ppt)\dotfill       & $4.7^{+1.3}_{-1.0}$ & $4.7^{+1.3}_{-1.0}$ & ... & ... \\
					~~~~$\delta_{\rm i'}$\dotfill   & Transit depth in $i'$ (ppt)\dotfill       & $4.48^{+1.2}_{-0.95}$ & $4.47^{+1.2}_{-0.97}$ & $3.32^{+0.63}_{-0.48}$ & $3.30^{+0.63}_{-0.47}$ \\
					~~~~$\delta_{\rm r'}$\dotfill   & Transit depth in $r'$ (ppt)\dotfill       & $5.0^{+1.4}_{-1.1}$   & $5.0^{+1.4}_{-1.1}$ & $3.76^{+0.72}_{-0.54}$ & $3.70^{+0.74}_{-0.56}$ \\
					~~~~$\delta_{\rm z'}$\dotfill   & Transit depth in $z'$ (ppt)\dotfill       & $4.16^{+1.1}_{-0.89}$ & $4.16^{+1.1}_{-0.90}$ & $3.16^{+0.60}_{-0.45}$ & $3.16^{+0.60}_{-0.44}$ \\
					~~~~$\delta_{\rm TESS}$\dotfill & Transit depth in TESS (ppt)\dotfill     & $4.22_{-0.63}^{+1.5}$               & $4.39^{+1.2}_{-0.95}$ & $3.28^{+0.61}_{-0.47}$ & $3.27^{+0.61}_{-0.46}$ \\
					~~~~$\tau$\dotfill     & Ingress/egress transit duration (days)\dotfill    & $0.00444^{+0.00062}_{-0.00054}$     & $0.00442^{+0.00077}_{-0.00058}$ & $0.00469^{+0.00052}_{-0.00045}$ & $0.00457^{+0.0021}_{-0.00087}$ \\
					~~~~$T_{14}$\dotfill   & Total transit duration (days)\dotfill             & $0.06913^{+0.00081}_{-0.00080}$    & $0.06904^{+0.00095}_{-0.00088}$ & $0.0728^{+0.0016}_{-0.0018}$ & $0.0728\pm0.0023$ \\
					~~~~$b$\dotfill        & Transit impact parameter\dotfill                  & $0.306^{+0.078}_{-0.12}$           & $0.26^{+0.18}_{-0.17}$ & $0.464^{+0.057}_{-0.071}$ & $0.44^{+0.21}_{-0.27}$ \\
					~~~~$\fave$\dotfill    & Incident Flux ($S_\oplus$)\dotfill                  & $1.7705^{+0.1616}_{-0.1175}$    & $1.6823^{+0.1689}_{-0.3673}$ & $2.4170\pm0.0881$ & $2.2554^{+0.1689}_{-0.6024}$ \\
					~~~~$d/R_*$\dotfill    & Separation at mid-transit\dotfill                 & $51.1^{+1.5}_{-1.4}$               & $48.2^{+4.0}_{-7.7}$ & $43.3\pm1.1$ & $40.5^{+6.0}_{-7.8}$ \\
					\smallskip\\\multicolumn{2}{l}{Predicted Parameters:}                      & \smallskip & & \smallskip & \\
					
					~~~~$M_P$\dotfill      & Planet mass$^{b}$ (\me)\dotfill                   & $3.8^{+1.6}_{-1.0}$                & $3.8^{+1.6}_{-1.0}$ & $4.3^{+1.7}_{-1.1}$                & $4.3^{+1.7}_{-1.1}$ \\
					~~~~$K$\dotfill        & RV semi-amplitude (m/s)\dotfill                   & $3.10^{+1.3}_{-0.85}$              & $3.4^{+1.8}_{-1.0}$ & $3.02^{+1.2}_{-0.76}$              & $3.40^{+1.8}_{-0.96}$ \\
					~~~~TSM\dotfill        & Transmission spectroscopy metric\dotfill          & 60.0                             &   50.0 & 38.0                               & 40.0 \\
					~~~~$M_P/M_*$\dotfill  & Planet-to-star mass ratio ($\times 10^{-5}$)\dotfill & $5.4^{+2.3}_{-1.5}$ & $5.5^{+2.4}_{-1.5}$ & $4.8^{+1.9}_{-1.2} $            & $4.9^{+1.9}_{-1.2}$ \\
					~~~~$\rho_P$\dotfill   & Planet mean density (cgs)\dotfill                 & $4.8^{+1.9}_{-1.3}$                & $4.7^{+1.9}_{-1.2}$ & $4.3^{+1.6}_{-1.0}$                & $4.3^{+1.6}_{-1.0}$ \\
					\hline
				\end{tabular}
			}%
		\end{center}
		$^{a}$The equilibrium temperature is estimated supposing a null-albedo and a 100\% efficient heat recirculation. $^{b}$The planetary mass is estimated using the \cite{Chen_Kipping2017} mass--radius relation. The radial velocity semi-amplitude ($K$) is predicted using the estimated mass. \label{fit:results_combined}
\end{table*}}
\section{Results and discussion} \label{discus}

\subsection{Radius valley}
\label{sec:RadVal}

The radius valley, discovered by \cite{Fulton_2017} from 1.5 to 2 \re\ for solar-type stars and by \cite{Cloutier_Menour_2020} from 1.4 to 1.7 \re\ for low-mass stars (i.e., $M \lesssim 0.65$ \msun), has attracted attention because it is important to understand planet formation and evolution. There are several theories to explain the origin of the radius valley: (1) thermally-driven mass loss, where planets lose their atmospheric envelopes because of stellar high energetic X-ray and ultra-violet (UV) radiation \citep{Lopez_2013, Owen_2013, Jin_2014, Chen_2016} and of the energy emerging from the cooling planetary cores \citep[core-powered mass loss,][]{Owen_2013,Ginzburg_2016,Ginzburg_2018}; (2) gas-poor formation, where some low-mass planetary cores accrete a thin gaseous envelope \citep[see e.g.][]{Lee_2021,Cherubim2023}. These planets can still experience thermally-driven mass loss; and (3) gas-depleted formation, where the population of super-Earths is formed late in a gas-depleted environment \citep[see e.g.][]{LopezRice2018,Cloutier_Menour_2020,Cherubim2023}; (4) Other studies based on coupled formation and evolution models predict that the sub-Neptunes are rich in water formed outside the ice-line and migrate toward the host stars \citep[][]{Venturini2020,Burn2024,Venturini2024}.

\begin{figure}[h!]
\centering
\includegraphics[width=\columnwidth]{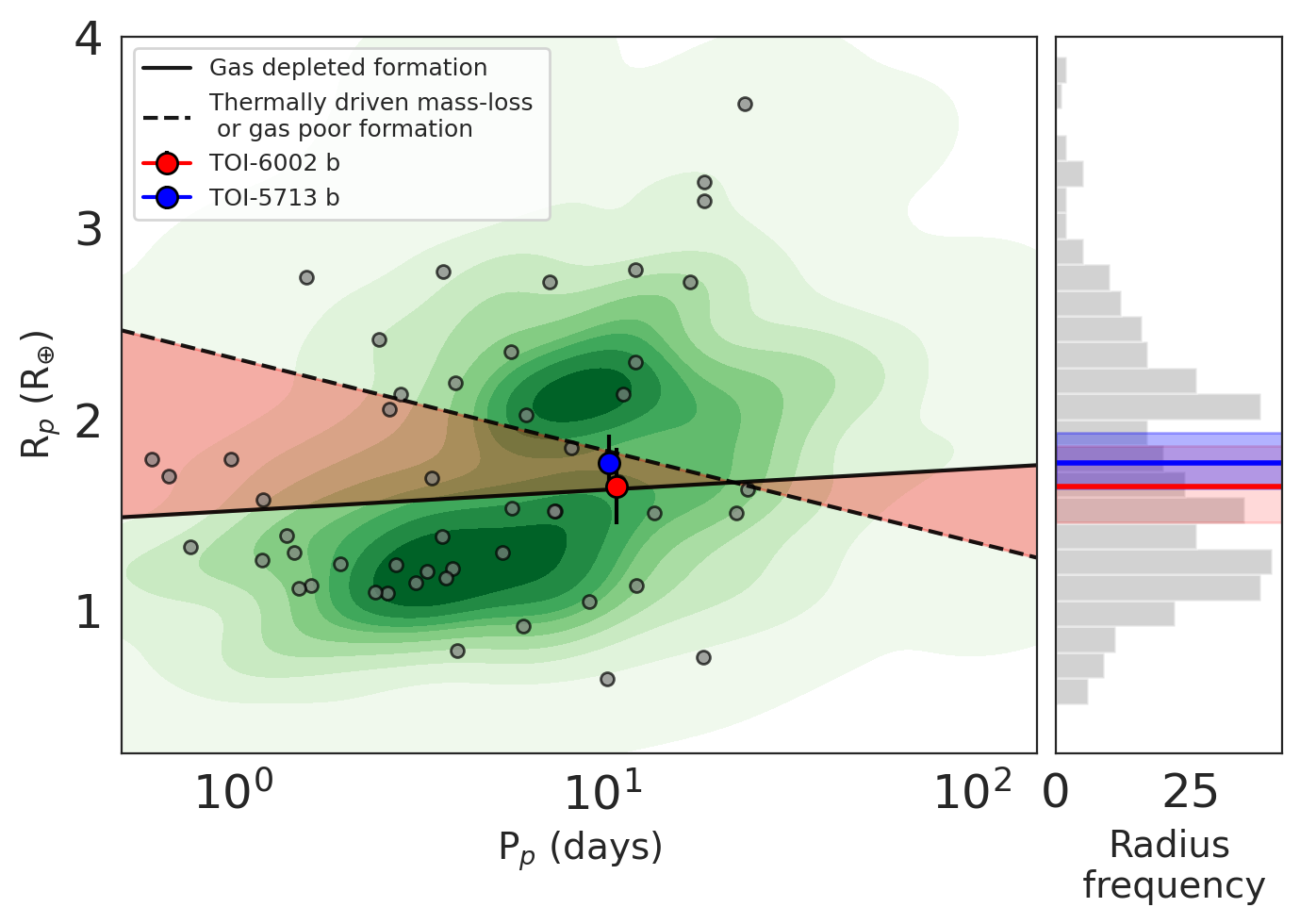}
\caption{Distribution of planet radii and orbital periods for all confirmed small planets hosted by low-mass stars (M$_{*} \lesssim 0.65 \msun$). The solid line represents the predicted location of the radius valley based on the gas-depleted formation model \citep{Cloutier_Menour_2020}. The dashed line shows the predicted location of the valley for the thermally-driven photoevaporation and gas-poor formation models \citep{VanEylen_2018}. The red and blue dots represents the planets TOI-6002b and TOI-5713b, respectively. The 1D radius distribution, with the location of the two planets, is shown on the right panel.}
\label{fig:RadVal}
\end{figure}

Observational and theoretical studies showed that the position of the radius valley is period-dependent (see Fig.~\ref{fig:RadVal}). Its location has been found trending to smaller planet radii for long periods in both thermally-driven mass loss and gas-poor formation models for solar FGK stars \citep[see e.g.][]{Fulton_2017,VanEylen_2018,LopezRice2018,Lee_2021}. For M dwarfs, its location has been found trending to larger radii for long periods in the gas-depleted theory \citep[see e.g][]{LopezRice2018,Cloutier_Menour_2020}.
Conversely, other studies on small planets around M dwarfs found a negative slope \citep[see e.g.][]{VanEylen2021, Bonfanti2024}, something similar to that of FGK stars. 

As shown in Fig.~\ref{fig:RadVal}, a wedge of opposing predictions emerged because of the difference in the slopes of the radius valley. Planets in this region are known as "keystone planets." They are predicted to be purely rocky by thermally-driven mass loss and gas-poor formation, and potentially rocky with small gaseous envelopes by the gas-depleted formation model \citep[see e.g.][]{Cherubim2023}. Only a few planets with measured densities are in this wedge. Having a large sample of keystone planets is important to understand the models dominating the formation and evolution of small planets. TOI-5713 b ($R_p=1.75_{-0.16}^{+0.14}$ \re\ and $P = 10.441989^{+0.000015}_{-0.000014}$\,d) is located within the wedge of keystone planets with the possibility of being at its upper edge, given the uncertainty on the planetary radius. TOI-6002 b ($R_p=1.63^{+0.21}_{-0.19}$\re\ and $P = 10.904821^{+0.000021}_{-0.000018}$\,d) is located at the lower edge of this wedge with the possibility of being inside it, given the uncertainty on the planetary radius. When their masses are determined with RV observations (see Sect.\ref{RV_}), these planets will join the small sample of keystone planets with density measurements. This sample will, in the future, help disentangle which model is dominating the formation of small (i.e., $R_p \lesssim 4 \re$) planets. 

In addition, for M dwarfs, a later study of \cite{Luque&palle2022} on small planets (with mass and radius precision better than 25\% and 8\%, respectively)  identified instead a density valley separating rocky planets, with density peaked at $\rho=0.94\pm0.13~\rho_{\oplus}$, and water-rich worlds, with density peaked at $\rho=0.47\pm0.05~\rho_{\oplus}$. A third peak was identified at $\rho=0.24\pm0.04~\rho_{\oplus}$, attributed to gas-rich planets. This study favored the pebble accretion model as the main mechanism that forms these populations of planets. In this model, water-rich planets are formed outside the ice-line and migrate toward the host stars. This finding was theoretically predicted by \cite{Venturini2020}, and newly supported by \cite{Burn2024} and \cite{Venturini2024} using coupled formation and evolution models. Conversely, \cite{Parviainen2024} stand against the existence of distinct population of water-rich planets around M dwarf stars. However, in all previous studies on radius and density valleys, strong conclusions require larger sample of precisely characterized exoplanets.

Given their predicted masses estimated from their radii using the relations of \cite{Chen_Kipping2017}, these planets are predicted to have densities of $0.855^{+0.345}_{-0.218}~\rho_{\oplus}$ for TOI-6002 b, and $0.782^{+0.2911}_{-0.182}~\rho_{\oplus}$ for TOI-5713 b. We also used the recent \texttt{SPRIGHT}\footnote{\url{https://github.com/hpparvi/spright}} python package of \cite{Parviainen2024}, where we found median values of the masses (and then densities) consistent with those obtained using \cite{Chen_Kipping2017} relations. To further guess their compositions, we referred to the radius space of the study of \citet{Luque&palle2022} and the corresponding density regimes in the density space. Given their radii, TOI-5713 b falls slightly towards the range of water-rich planets, and TOI-6002 b falls in the region where we find both rocky and water-rich planets. Therefore, it is uncertain whether the two planets could be rocky or water-rich worlds. However, even if they are not of intermediate composition, they might then be either some of the largest rocky planets or some of the smallest water-rich worlds known today. As for the radius valley, both planets straddle near the opposing side of the region of "keystone planets," where it is difficult to guess whether they lost their gaseous envelopes or not. When their masses are determined, these two planets can be added to the sample of "keystone planets," which will be used to conduct in-depth statistical inferences on the radius and density valleys. Such studies will help to refine the relative dominance of the various mechanisms proposed to be responsible for the formation and evolution of super-Earths and sub-Neptunes around M-dwarf stars. 

\subsection{Prospects for a radial velocity follow-up} \label{RV_}

In Table~\ref{fit:results_combined}, we show the predicted semi-amplitudes for TOI-6002\,b  and TOI-5713\,b for both circular and eccentric orbits. With both planets inducing reflex motions to their host-stars with semi-amplitudes on the order of 3\,${\rm m s^{-1}}$, high-resolution, well-stabilized spectrographs are needed for their measurement. 
Due to the relatively faint apparent magnitudes of both stars, reaching such precision will be challenging for instruments on 4-m class telescopes, such as HARPS-N \citep{Cosentino12} in the optical. This is also true for NIR instruments such as CARMENES \citep{Quirrenbach_2020S} or SPIRou \citep{Donati18} which would allow to reach a RV precision in the order of ~10\,${\rm m s^{-1}}$ or less.
The MAROON-X instrument \citep{2020SPIE11447E..1FS} at the 8.1-m Gemini North telescope offers the necessary stability and throughput to successfully measure the reflex motions of both stars. Assuming exposure times of 900\,s, the instrument's red arm will achieve a maximum signal-to-noise-ratio larger than 95 and 65 for TOI-6002 and TOI-5713, respectively. 

Following \cite{Engle:2023}, we can expect negligible line broadening and a low activity level for TOI-6002 due to the age of this star. Assuming an additional noise term of 1\,${\rm m s^{-1}}$ accounting for stellar activity, we expect that the reflex motion can be detected at a 5$\sigma$ level with less than 11h telescope time using MAROON-X. This would not only allow us to measure the mass of TOI-6002~b, but also to derive the eccentricity of the planet's orbit.

In Sect.~\ref{sec:activity}, we determine the stellar rotation period of TOI-5713 from photometry. This translates to a maximum rotational line broadening ($v\,\sin i_{*}$) on the order of $1.6\,{\rm km\,s^{-1}}$, which is below the resolution limit for most high-resolution spectrographs, and thus does not affect the radial-velocity precision. Nevertheless, we observe moderate activity and a photometric rotational variability close to the orbital period of TOI-5713~b, making activity mitigation challenging. Assuming an optimistic noise term of 3\,${\rm m s^{-1}}$ accounting for this stellar activity after successful mitigation, a 5$\sigma$ detection of the radial velocity signal due to TOI-5713\,b would need at least 16\,hr telescope time with MAROON-X.

\subsection{Potential for atmospheric characterization}

TOI-6002b and TOI-5713b, with radii of 1.63 R$_{\oplus}$ and 1.75 R$_{\oplus}$ and incident stellar fluxes of 1.77 and 2.42 times that of Earth respectively, could be classified as 'super-Venus' planets if they have rocky compositions. Venus, in comparison, receives about 1.9 times the flux of Earth and is 95\% the size of Earth. According to \citet{kane2014}, Venus analogs are defined as predominantly rocky planets within the "Venus-Zone (VZ)," receiving insolation fluxes between approximately 0.95 and 25 times that of Earth. This classification suggests that both TOI-6002b and TOI-5713b fall within this category. 
With incident flux slightly lower (TOI-6002b) and slightly higher (TOI-5713b) than modern Venus, these two exoplanets are intriguing candidates to expand our understanding of the evolution of rocky planets in the state from a temperate Earth-like to a hot Venus-like planet. The incident flux levels indicate that both planets might experience or might have experienced strong runaway greenhouse effects, depending on factors like the initial water inventory, planetary evolution and atmospheric escape rates \citep{kaltenegger2023hot}. Combined with the estimated ages of the stars, this could indicate atmospheric conditions similar to modern Venus. However, studying planets with incident irradiation higher than modern Earth to modern Venus levels and beyond, that may be experiencing strong greenhouse effects or may have evolved into a post-runaway greenhouse state is crucial, as it can provide insights into the factors that cause some rocky planets to resemble Venus-analogs while others may maintain Earth-analog characteristics \citep{ehrenreich2012venus, ch-surface, jordan2021photochemistry, kaltenegger2023hot}.

To investigate two cases of potential atmospheric characteristics of these planets, we constructed models simulating a) cloudless and 10 mbar cloudy, 92-bar Venus-like, and b) 1-bar Earth-like atmospheres. 
We link the one-dimensional VULCAN chemical kinetic code \citep{vulcan} to the one-dimensional radiative transfer model PETITRADTRANS to predict the transmission spectra for these clear atmospheres \citep{mol}. The radiative--convective temperature--pressure profiles were computed with the HELIOS code \citep{malik}. The species considered for our atmospheric models include N$_{2}$, O$_{2}$, O$_{3}$ H$_{2}$O, CO$_{2}$, CO, HCN, CH$_{4}$, NO, NH$_{3}$, and C$_{2}$H$_{2}$ for the Earth-like atmosphere (rich in N$_{2}$). Additionally, Rayleigh scattering for N$_{2}$ and O$_{2}$, as well as collision-induced absorption for the N$_{2}$--N$_{2}$ and O$_{2}$--O$_{2}$ pairs, are considered. For the Venus-like atmosphere (rich in CO$_{2}$), we include H$_{2}$S, SO$_{2}$, OCS, H$_{2}$O, CO$_{2}$, CO, CH$_{4}$, HCN, and C$_{2}$H$_{2}$ as well as CO$_{2}$--CO$_{2}$ collision-induced absorption, as opacity sources in the atmosphere. Rayleigh scattering is included for CO$_{2}$.

\citet{ehrenreich2012venus} showed that during Venus's 2012 transit, its transmission spectra indicated that the cloud and haze layers obstruct probing below an altitude of 80\,km. \citet{pidhorodetska202198} modeled the transmission spectra of the L~98-59 planets, a benchmark system for studying the VZ planets \citep{ostberg2023demographics}. They showed that if the L~98-59 planets possess a clear Venus-like atmosphere, NIRSpec could detect CO$_2$ within 26 transits for each planet. However, the presence of H$_2$SO$_4$ clouds would significantly suppress CO$_2$ absorption. 
A new study by \citet{kaltenegger2023hot} shows that there is a wide range of possible atmospheres for a rocky planet experiencing a runaway greenhouse effect, which changes the atmosphere and observable features significantly. Model transmission spectra of the planet LP~890-9c \citep{Delrez2022}, another benchmark system for studying the VZ planets \citep{kaltenegger2023hot} show that depending on the state of the planet, {\it James Webb} Space Telescope (\JWST) observations could infer evidence of H$_2$O with 3 transit (at 3$\sigma$ confidence) for a full runaway greenhouse scenario for LP~890-9c, and CO$_2$-dominated atmospheres resembling Venus without high-altitude terminator clouds with 8 transits 
\citep{Barrientos2023}. A 3D Venus model for the same planet with higher, thick cloud coverage increases the amount of transits needed to be able to explore the atmospheric composition. Note that all predictions could be complicated by the impact of clouds and (or) unocculted starspots, especially for cool host stars. 

These results show that observing these exoplanets is of critical importance to understanding their evolutionary stage and the effect of increasing solar irradiation on rocky worlds. Additionally, planetary systems like the TRAPPIST-1 system could also allow insight into the effect of the increase of irradiation on the environment on a planet (e.g. \citet{Payne2024}). However, even for a specific orbital distance and stellar irradiation, the observable spectral features in transit for an exoplanet depend on their specific atmospheric evolution and cloud coverage (e.g., \citet{Kaltenegger2017, Kaltenegger2020}), requiring a wider range of observations to explore the characteristic changes of rocky worlds with stellar irradiation. TOI-6002b and TOI-5713b provide two important datapoints in that exploration. 

\begin{figure*}
    \centering
    \includegraphics[width=0.5\linewidth]{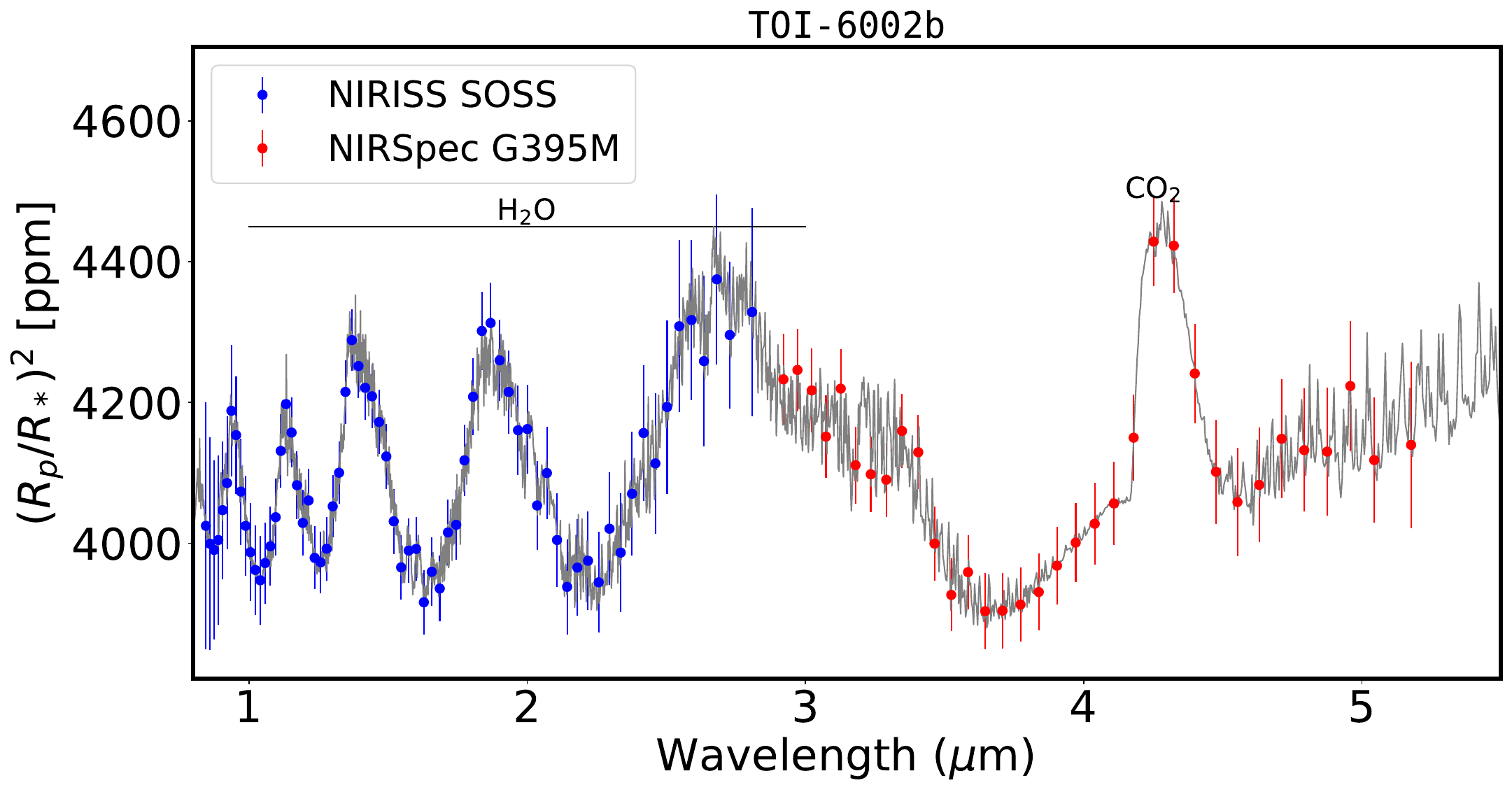}\hfill
    \includegraphics[width=0.5\linewidth]{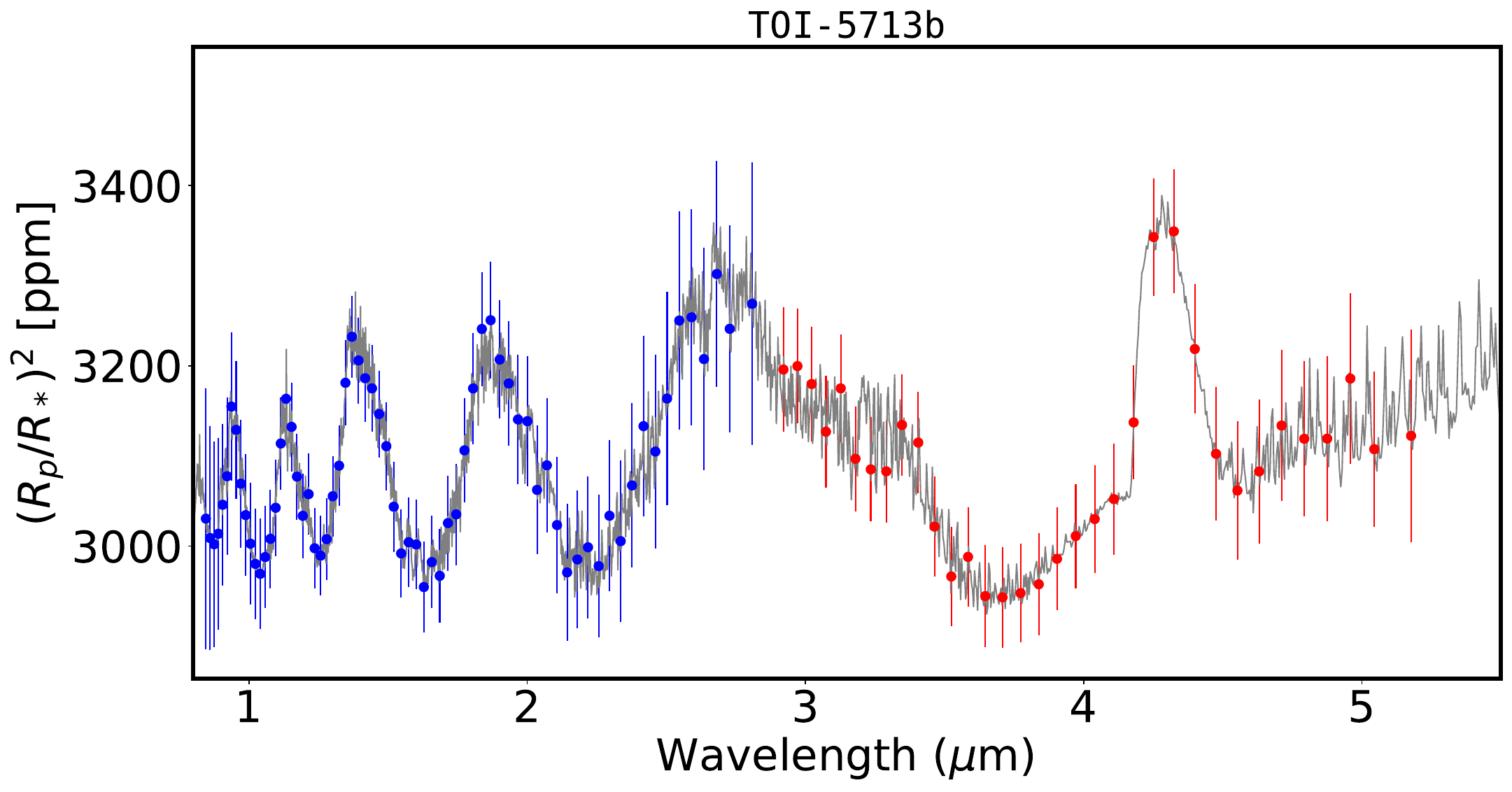}\hfill
    \includegraphics[width=0.5\linewidth]{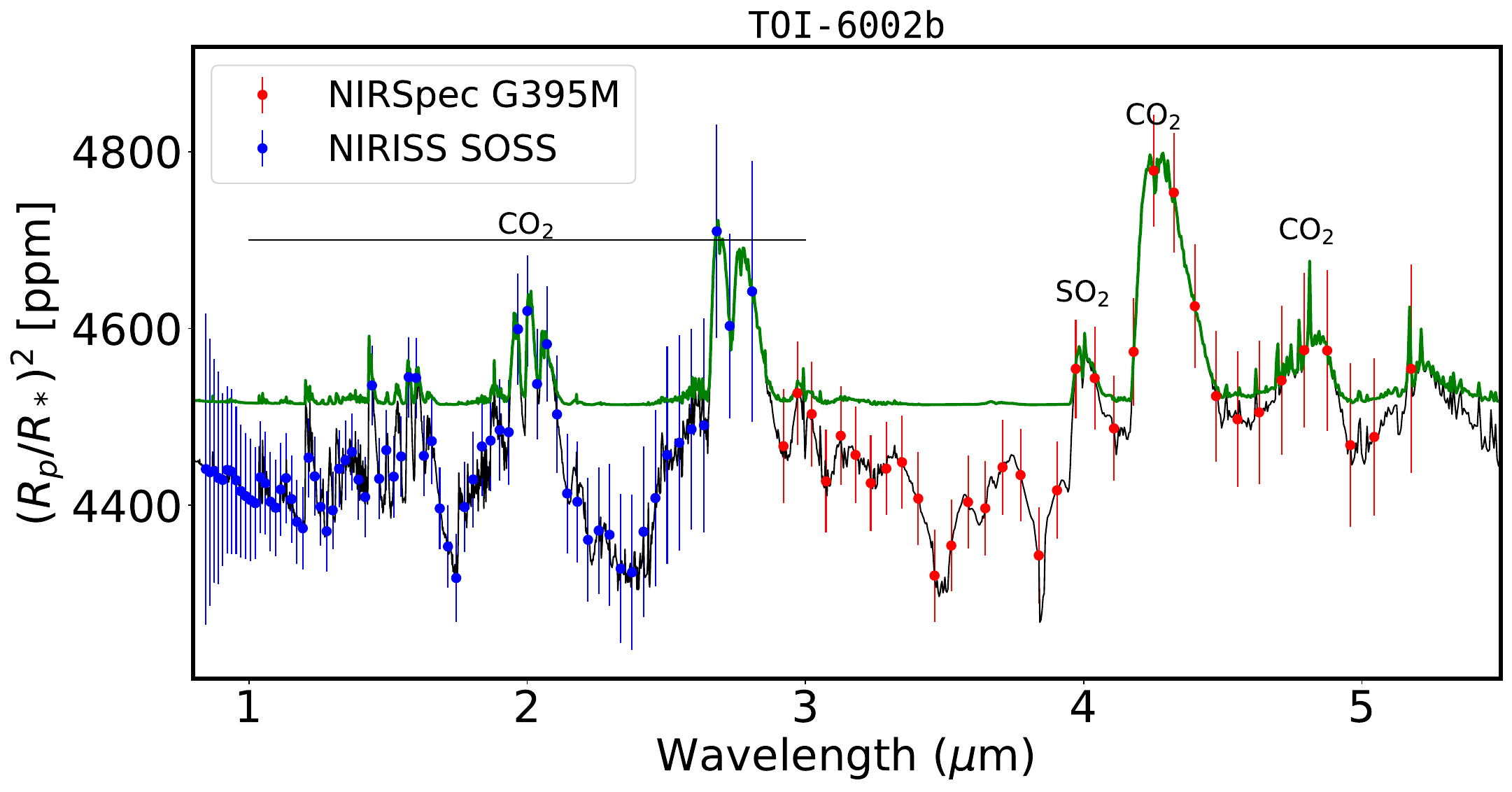}\hfill
    \includegraphics[width=0.5\linewidth]{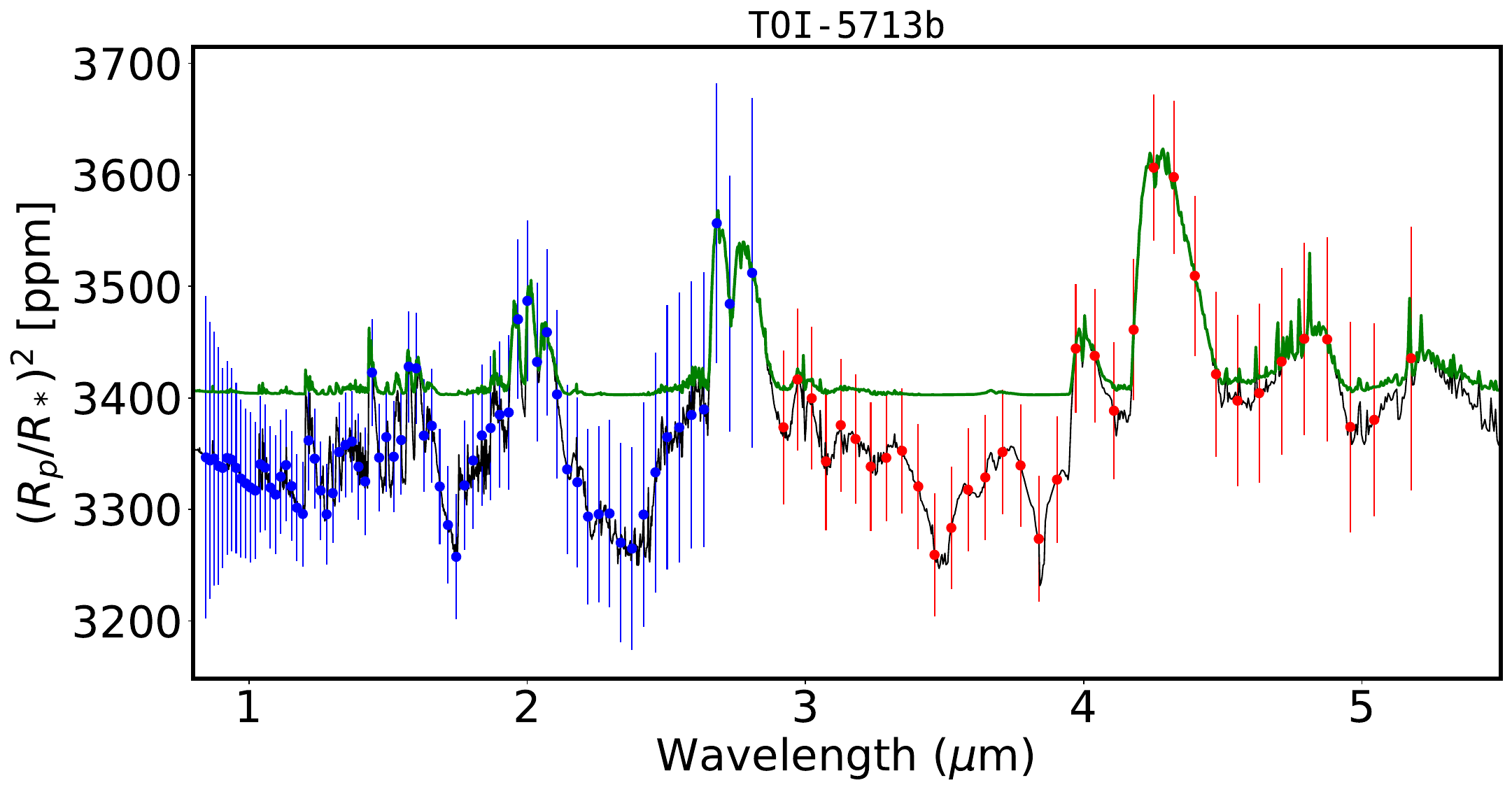}
    \caption{Simulated \JWST\ transmission spectra for TOI-6002b and TOI-5713b. The top panel shows spectra for a cloudless, 1-bar Earth-like atmosphere. The bottom panel displays spectra for a 92-bar Venus-like atmosphere, both cloudless (black) and with a cloud deck at 10 mbar (green). Error bars are shown for a transit observation simulated at R = 30.}
    \label{tr_sp}
\end{figure*}

For TOI-6002b and TOI-5713b, the synthetic transmission spectra generated from our models (Earth-like surface temperatures are 524 K and 580 K, respectively; Venus-like surface temperatures are 1053 K and 1110 K, respectively) show key differences between Earth-like and Venus-like atmospheres (see Fig.~\ref{tr_sp}). 

Models of TOI-6002b and TOI-5713b for clear hot Earth-like atmospheric compositions are dominated by water vapor. Water vapor is a strong absorber across the infrared for Earth-like atmospheric compositions, with amplitudes of about 300 ppm in the mid-infrared (see Fig.~\ref{tr_sp}, top panel). Note that Fig.~\ref{tr_sp} (top panel) also shows a notable CO$_2$ feature at 4--5\,$\mu$ with a depth of 400 ppm. However, we would assume the atmospheric CO$_2$ concentration to decrease during a runaway greenhouse stage, reducing that feature strength. Because we are not modeling the possible evolution of the planet's atmosphere, we have not changed the CO$_2$ concentration here, but note that the features could be much smaller than shown.

Models of TOI-6002b and TOI-5713b using clear Venus-like atmospheric composition are primarily characterized by CO$_2$ opacity, with a depth of about 250 ppm in the mid-infrared, and a prominent SO$_2$ feature at 4 $\mu$ and between 7 and 10\,$\mu$, with depths above $\approx$ 100 ppm. With a cloud deck at 10 mbar, the CO$_2$ feature at approximately 4.2 $\mu$m remains discernible, exhibiting a depth greater than 200 ppm.

These models show that TOI-6002b and TOI-5713b represent promising targets for atmospheric characterization with \JWST, as they may provide valuable insights into the diverse evolutionary pathways and climate states of hot rocky or water-rich exoplanets. Mass measurements could help provide better constraints on the viability of atmospheric studies and would be possible with the MAROON-X instrument at the 8.1-m Gemini North telescope.   

\section{Conclusion} \label{concl}

In this paper, we reported the TESS discovery and initial characterization of TOI-6002 b and TOI-5713 b, two super-Earth planets orbiting the mid-M-dwarf stars TOI-6002 (V = 14.6) and TOI-5713 (V = 15.355). We used the 2-min-cadence TESS observations from four sectors for each target, ground-based photometry, high-angular resolution imaging, and spectroscopic observations to validate the planetary nature of the detected transit events. We jointly analyzed the transit light curves observed by TESS and ground-based facilities to characterize the planets. 

TOI-6002 b has a radius of $1.65^{+0.22}_{-0.19} \re$, an orbital period of $10.904821^{+0.000021}_{-0.000018}$ days, and receives $1.77^{+0.16}_{-0.11} S_\oplus$. TOI-5713 b has a radius of $1.77_{-0.11}^{+0.13} \re$, an orbital period of $10.441989^{+0.000015}_{-0.000014}$ days, and receives $2.42\pm{0.11} S_\oplus$. Both planets are located just outside but near the inner edge of the habitable zone around their host stars, making them interesting targets for future studies to explore the evolution of exoplanets from hot, potentially habitable to Venus-like worlds. These two planets provide compelling additions to the sample of planets critical for understanding the formation and evolution of small worlds on the basis of the radius and density valleys. 

\begin{acknowledgements}
This research was carried out in part at the Jet Propulsion Laboratory, California Institute of Technology, under a contract with the National Aeronautics and Space Administration (80NM0018D0004).
This article is based on observations made with the MuSCAT2 instrument, developed by ABC, at Telescopio Carlos S\'anchez operated on the island of Tenerife by the IAC in the Spanish Observatorio del Teide.
TRAPPIST-North is a project funded by the University of Liege (Belgium), in collaboration with Cadi Ayyad University of Marrakesh (Morocco).
We acknowledge the use of public TESS data from pipelines at the TESS Science Office and at the TESS Science Processing Operations Center.
This work makes use of observations from the LCOGT network. Part of the LCOGT telescope time was granted by NOIRLab through the Mid-Scale Innovations Program (MSIP). MSIP is funded by NSF.
This research has made use of the Exoplanet Follow-up Observation Program (ExoFOP; DOI: 10.26134/ExoFOP5) website, which is operated by the California Institute of Technology, under contract with the National Aeronautics and Space Administration under the Exoplanet Exploration Program.
We acknowledge the use of public TESS data from pipelines at the TESS Science Office and at the TESS Science Processing Operations Center.
This paper includes data collected by the TESS mission that are publicly available from the Mikulski Archive for Space Telescopes (MAST).
Funding for the TESS mission is provided by NASA's Science Mission Directorate. KAC acknowledges support from the TESS mission via subaward s3449 from MIT.
Resources supporting this work were provided by the NASA High-End Computing (HEC) Program through the NASA Advanced Supercomputing (NAS) Division at Ames Research Center for the production of the SPOC data products.
This paper is based on observations made with the MuSCAT3 instrument, developed by the Astrobiology Center and under financial supports by JSPS KAKENHI (JP18H05439) and JST PRESTO (JPMJPR1775), at Faulkes Telescope North on Maui, HI, operated by the Las Cumbres Observatory. 
This work was partially supported by a grant from the Erasmus+ International Credit Mobility program (M.~Ghachoui).
This publication benefits from the support of the French Community of Belgium in the context of the FRIA Doctoral Grant awarded to MT.
MG is F.R.S-FNRS Research Director.
The postdoctoral fellowship of KB is funded by F.R.S.-FNRS grant T.0109.20 and by the Francqui Foundation.
Author F.J.P acknowledges financial support from the Severo Ochoa grant CEX2021-001131-S funded by MCIN/AEI/10.13039/501100011033 and 
Ministerio de Ciencia e Innovación through the project PID2022-137241NB-C43. 
H.P. acknowledges support by the Spanish Ministry of Science and Innovation with the Ramon y Cajal fellowship number RYC2021-031798-I.
This work is partly supported by JSPS KAKENHI Grant Number JP18H05439 and JST CREST Grant Number JPMJCR1761.
E.E-B. acknowledges financial support from the European Union and the
State Agency of Investigation of the Spanish Ministry of Science and
Innovation (MICINN) under the grant PRE2020-093107 of the Pre-Doc
Program for the Training of Doctors (FPI-SO) through FSE funds
We acknowledge financial support from the Agencia Estatal de Investigaci\'on of the Ministerio de Ciencia e Innovaci\'on MCIN/AEI/10.13039/501100011033 and the ERDF “A way of making Europe” through project PID2021-125627OB-C32, and from the Centre of Excellence “Severo Ochoa” award to the Instituto de Astrofisica de Canarias.
This material is based upon work supported by the National Aeronautics and Space Administration under Agreement No.\ 80NSSC21K0593 for the program ``Alien Earths''.
The results reported herein benefited from collaborations and (or) information exchange within NASA’s Nexus for Exoplanet System Science (NExSS) research coordination network sponsored by NASA’s Science Mission Directorate.
This research was supported by Wallonia-Brussels International.
\end{acknowledgements}
\bibliographystyle{aa}
\bibliography{bib.bib}

clearpage

\onecolumn
\begin{appendix}

\section{Additional figures}

\begin{figure*}[hbt!]
    \centering
    \includegraphics[width=0.375\textwidth]{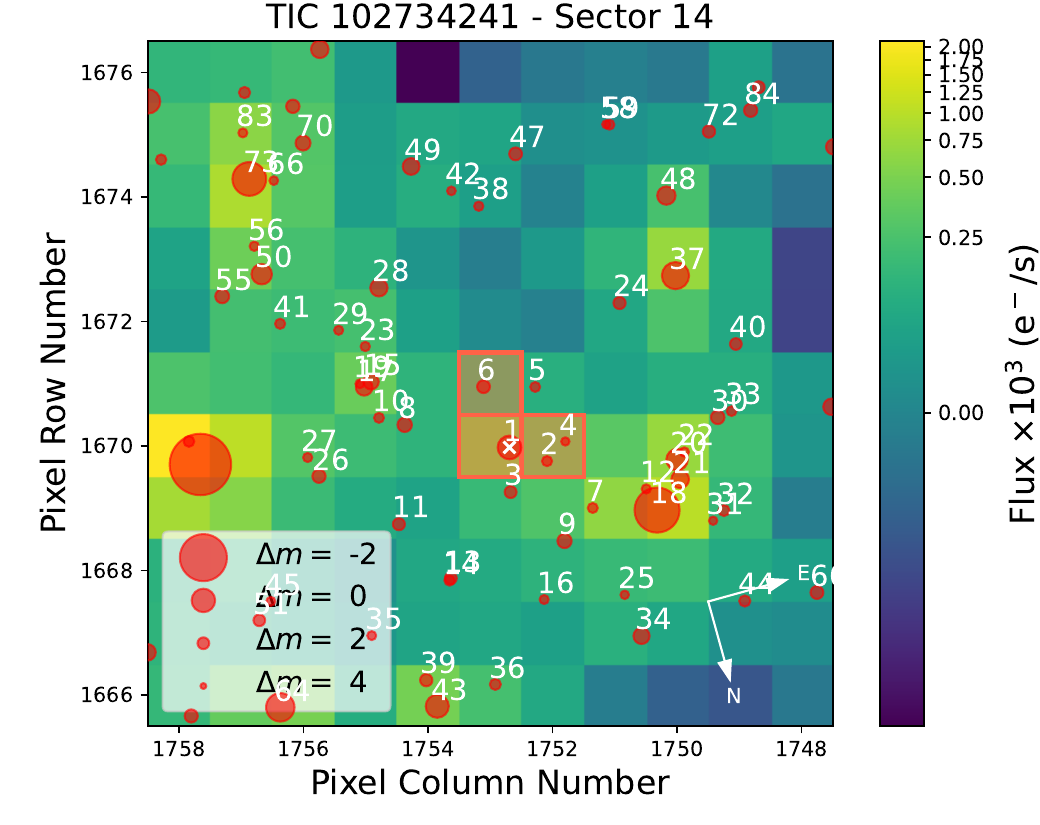}
    \includegraphics[width=0.375\textwidth]{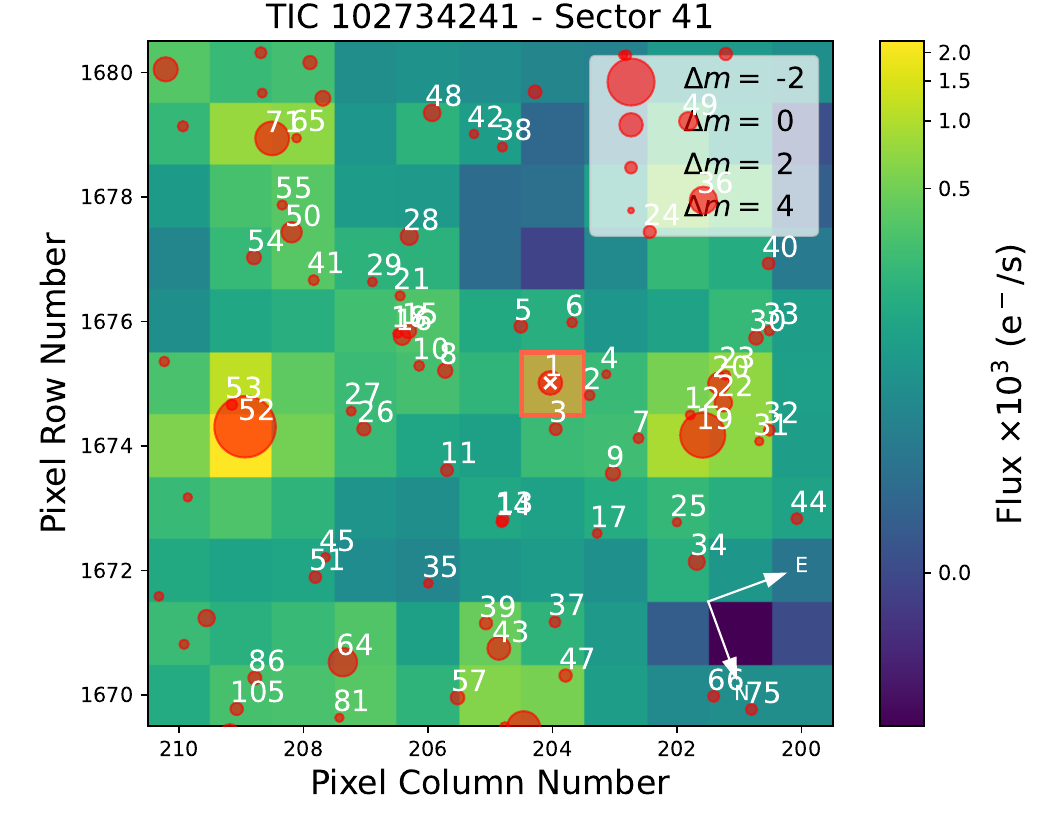}
    \includegraphics[width=0.375\textwidth]{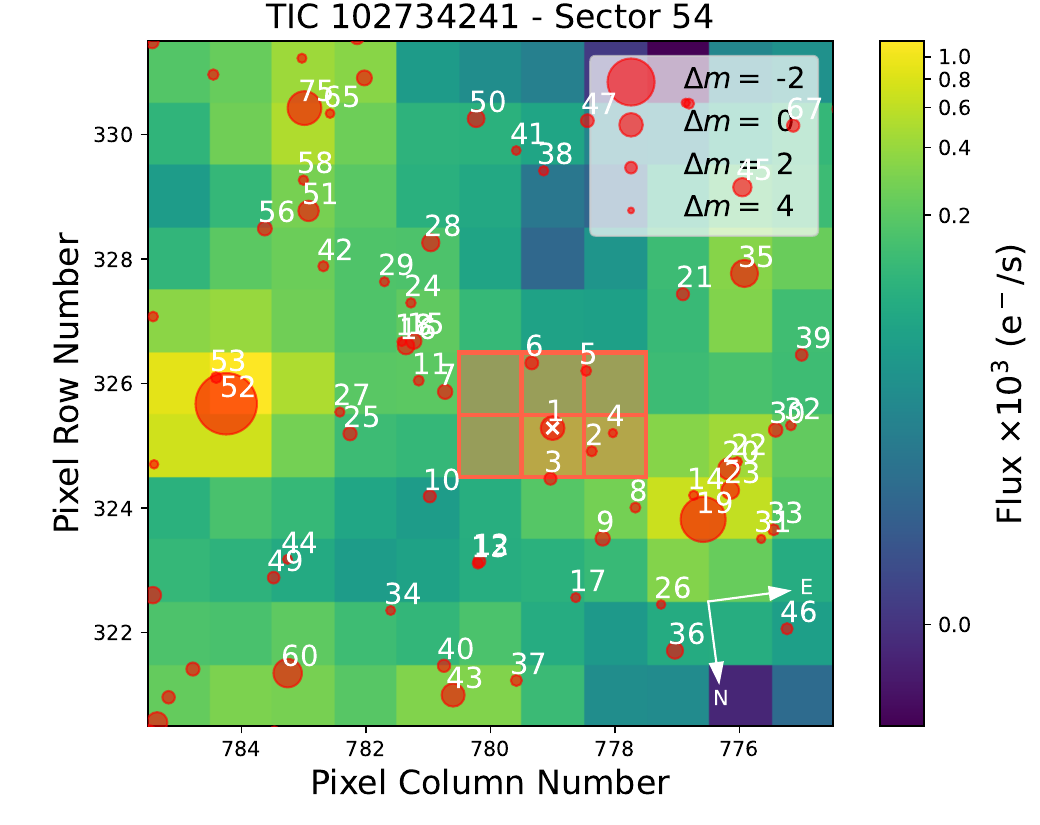}
    \includegraphics[width=0.375\textwidth]{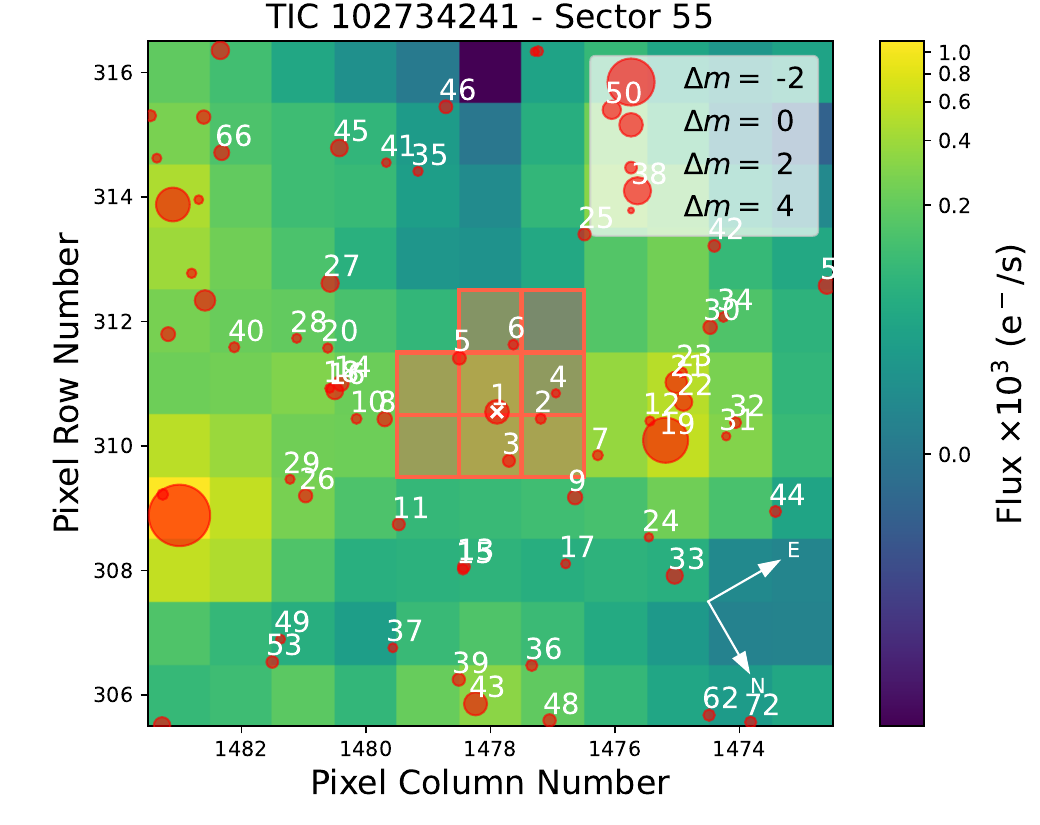}
    \includegraphics[width=0.375\textwidth]{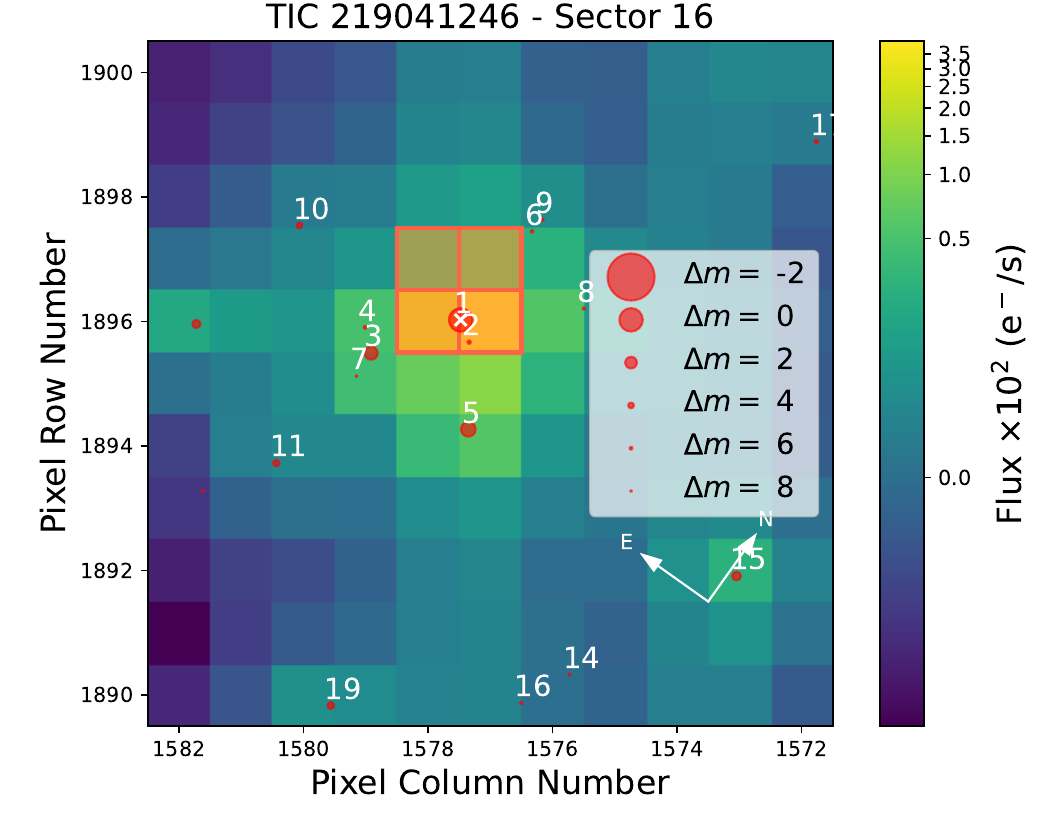}
    \includegraphics[width=0.375\textwidth]{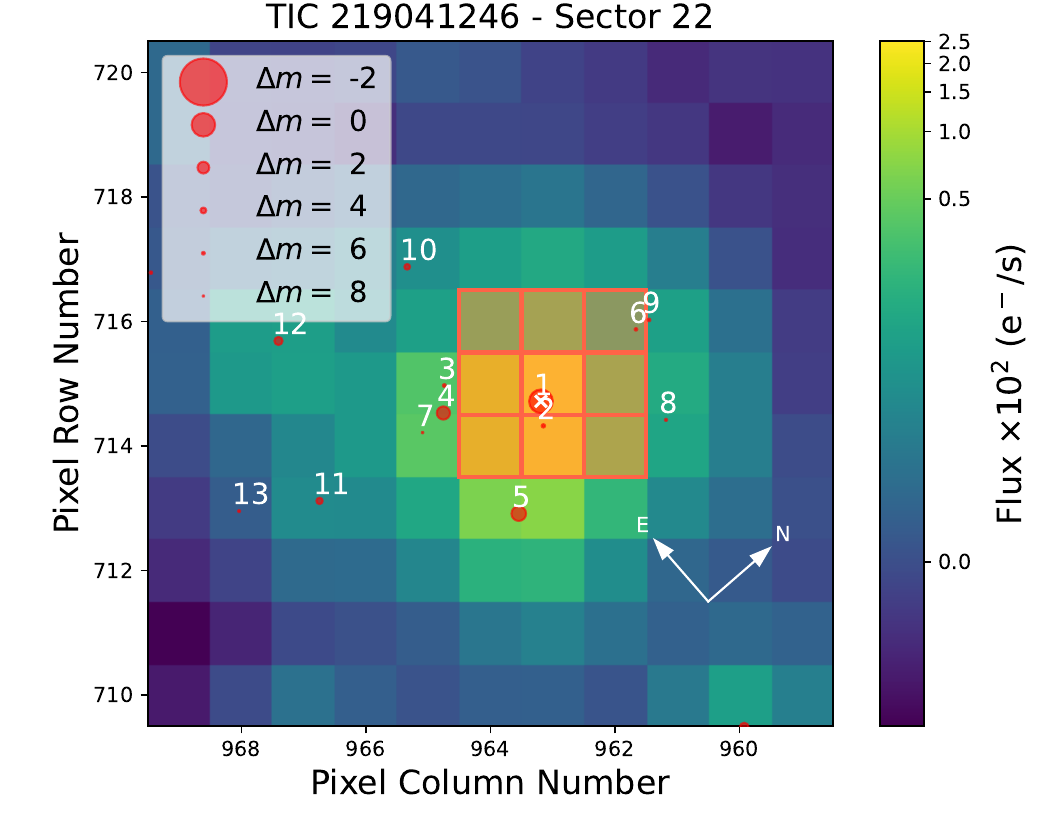}
    \includegraphics[width=0.375\textwidth]{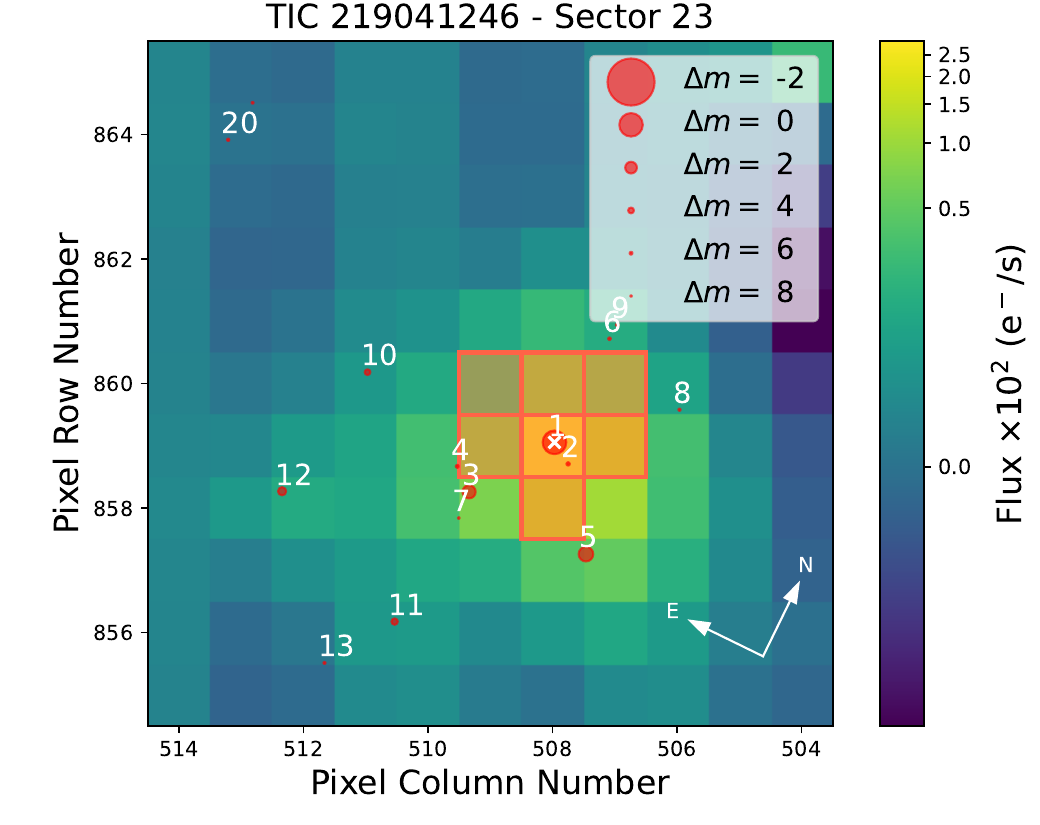}
    \includegraphics[width=0.375\textwidth]{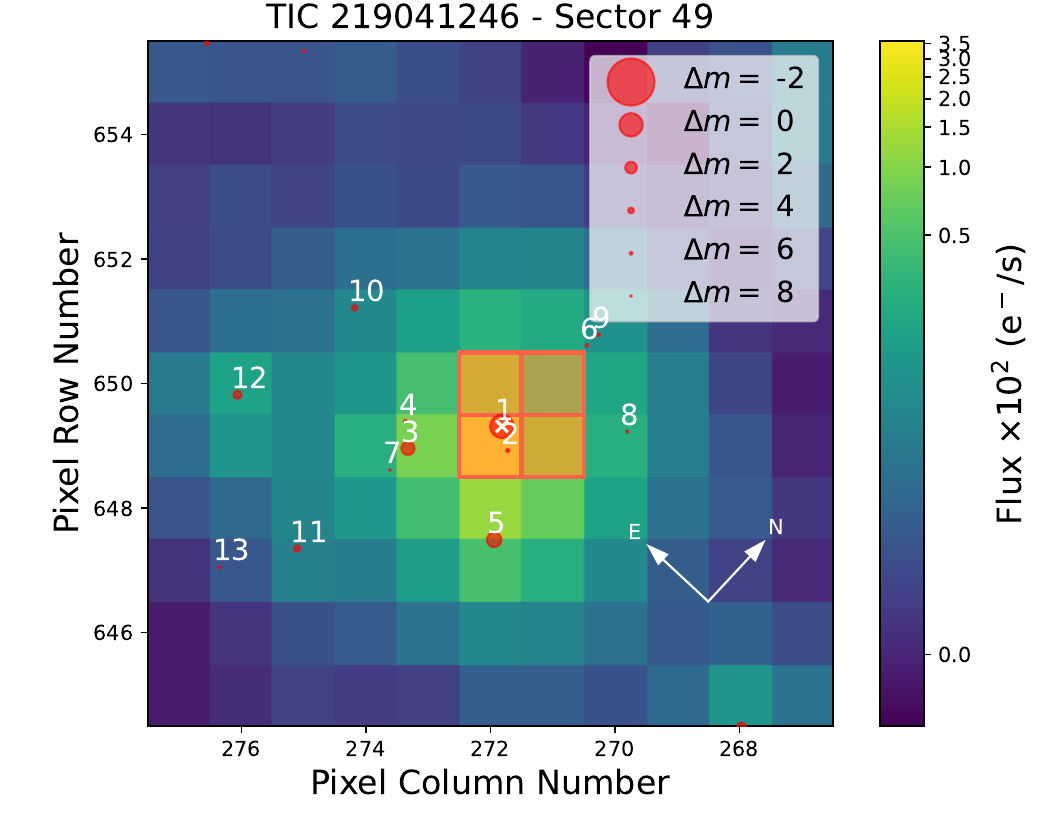}
    \caption{Target pixel files \textcolor{red}{}(TPFs) of TOI-6002 (TIC 102734241) in TESS sectors 14, 41, 54, 55, and TOI-5713 (TIC 219041246) in sectors 16, 22, 23, and 49 created with \texttt{tpfplotter} \citep{Aller_2020A&A}. Regions in orange highlight the aperture used to extract the SPOC photometry. Red circles represent the \gaia\ DR2 sources, with sizes representing the magnitude contrasts with respect to the targets.}
\label{TESS:fov}
\end{figure*}

\clearpage

\begin{figure*}
    \centering
    \includegraphics[width=\textwidth]{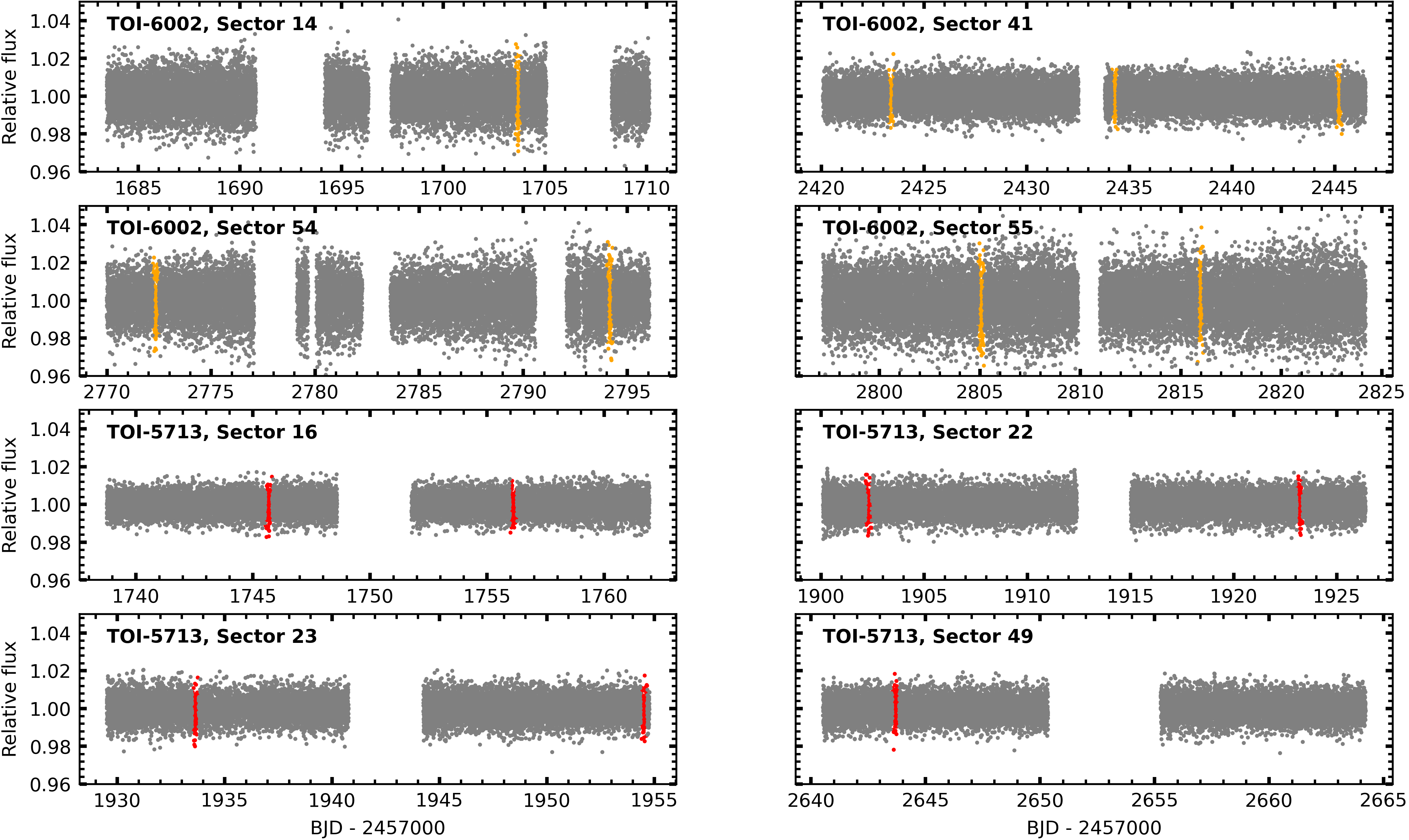}
    \caption{Detrended TESS photometry of TOI-6002\,b and TOI-5713\,b. The flares in the TOI-573 light curves were removed using an upper 3$\sigma$ clipping. Gray points show the detrended PDC-SAP fluxes from the SPOC pipeline. Orange and red points highlight the transits of TOI-6002\,b and TOI-5713\,b, respectively}.
\label{TESS:LC}
\end{figure*}

\end{appendix}

\end{document}